\documentclass[sigconf]{acmart}
\usepackage{hyperref}
\usepackage{enumitem}
\usepackage{xspace}
\usepackage{listings}
\usepackage{color}
\usepackage{soul}
\usepackage[noabbrev,capitalise]{cleveref}
\usepackage[linesnumbered,ruled,vlined]{algorithm2e}

\usepackage{tikz}
\newcommand*\circled[1]{\tikz[baseline=(char.base)]{
            \node[shape=circle,draw,inner sep=0.3pt] (char) {#1};}}
            
\usepackage{xcolor}
\newcommand*\circleddark[1]{\tikz[baseline=(char.base)]{
            \node[shape=circle,fill,inner sep=0.3pt] (char) {\textcolor{white}{#1}};}}

\definecolor{nav}{RGB}{0,0,128}

\usepackage{xcolor}

\copyrightyear{2021}
\acmYear{2021}
\setcopyright{acmlicensed}\acmConference[SIGMOD '21]{Proceedings of the 2021 International Conference on Management of Data}{June 20--25, 2021}{Virtual Event, China}
\acmBooktitle{Proceedings of the 2021 International Conference on Management of Data (SIGMOD '21), June 20--25, 2021, Virtual Event, China}
\acmPrice{15.00}
\acmDOI{10.1145/3448016.3457330}
\acmISBN{978-1-4503-8343-1/21/06}

\author{{Jinglin Peng$^{\dagger*}$ \hspace{.2em} Weiyuan Wu$^{\dagger*}$ \hspace{.2em} Brandon Lockhart$^{\dagger}$ \hspace{.2em} Song Bian$^{\ddagger}$ \hspace{.2em} Jing Nathan Yan$^{\diamondsuit}$ \hspace{.2em} Linghao Xu$^{\dagger}$ \hspace{.2em} Zhixuan Chi$^{\dagger}$ \hspace{.2em} Jeffrey M. Rzeszotarski$^{\diamondsuit}$ \hspace{.2em} Jiannan Wang$^{\dagger}$}}

\renewcommand{\authors}{Jinglin Peng, Weiyuan Wu, Brandon Lockhart, Song Bian, Jing Nathan Yan, Linghao Xu, Zhixuan Chi, Jeffrey M. Rzeszotarski, and Jiannan Wang}

\affiliation{
  \institution{Simon Fraser University$^{\dagger}$ \hspace{2em} Cornell University$^{\diamondsuit}$ \hspace{2em} The Chinese University of Hong Kong$^{\ddagger}$ }
  \country{}
}

\affiliation{\small
  {\{jinglin\_peng, youngw, brandon\_lockhart, linghaox, zhixuan\_chi, jnwang\}@sfu.ca}$^{\dagger}$ {\hspace{1em} \{jy858, jeffrz\}@cornell.edu}$^{\diamondsuit}$ \hspace{1em}  sbian@se.cuhk.edu.hk$^{\ddagger}$
  \country{}
}

\thanks{* Both authors contributed equally to this research.}
\thanks{\textbf{Acknowledgements.} This work was supported in part by Mitacs through an Accelerate Grant, NSERC through a discovery grant and a CRD
grant as well as NSF grant IIS-1850195. All opinions, findings, conclusions and recommendations in this paper are those of the
authors and do not necessarily reflect the views of the funding agencies.}

\newcommand{\stitle}[1]{\vspace{0.1em}\noindent{\bf #1}}
\newcommand{\msection}[1]{\section{#1}}
\newcommand{\msubsection}[1]{\vspace{-.5em}\subsection{#1}}
\newcommand{\edax}{\textsf{DataPrep.EDA}\xspace}
\newcommand{\pp}{\textsf{Pandas-profiling}\xspace}
\newcommand{\pandas}{\textsf{Pandas}\xspace}
\newcommand{\scikitlearn}{\textsf{Scikit-Learn}\xspace}
\newcommand{\numpy}{\textsf{NumPy}\xspace}
\newcommand{\pandasviz}{\textsf{Pandas+Plotting}\xspace}
\newcommand{\pyviz}{\textsf{Plotting}\xspace}
\newcommand{\matplotlib}{\textsf{Matplotlib}\xspace}
\newcommand{\bokeh}{\textsf{Bokeh}\xspace}

\newcommand{\dask}{\textsf{Dask}\xspace}
\newcommand{\seaborn}{\textsf{Seaborn}\xspace}

\newcommand{\add}[1]{{{#1}\xspace}}
\newcommand{\final}[1]{{{{#1}}\xspace}}

\begin{document}

\pagestyle{empty}
\fancyhead{}

\title{DataPrep.EDA: Task-Centric Exploratory Data Analysis \\ for Statistical Modeling in Python}

\begin{abstract}
Exploratory Data Analysis (EDA) is a crucial step in any data science project. However, existing Python libraries fall short in supporting data scientists to complete common EDA tasks for statistical modeling. Their API design is either too low level, which is optimized for plotting rather than EDA, or too high level, which is hard to specify more fine-grained EDA tasks. In response, we propose \edax, a novel task-centric EDA system in Python. \edax allows data scientists to declaratively specify a wide range of EDA tasks in different granularity with a single function call. We identify a number of challenges to implement \edax, and propose effective solutions to improve the scalability, usability, customizability of the system. In particular, we discuss some lessons learned from using \dask  to build the data processing pipelines for EDA tasks and describe our approaches to accelerate the pipelines. We conduct extensive experiments to compare \edax with \pp, the state-of-the-art EDA system in Python. The experiments show that \edax significantly outperforms \pp in terms of both speed and user experience. \edax is open-sourced as an EDA component of DataPrep: {\color{blue}\url{https://github.com/sfu-db/dataprep}}.
\end{abstract}

\maketitle

\vspace{-.5em}
\section{Introduction}

Python has grown to be one of the most popular programming languages in the world~\cite{tiobe-index} and is widely adopted in the data science community. For example, the Python data science ecosystem, called PyData, is used by universities and online learning platforms to teach data science essentials~\cite{ucb-ds,ibm-ds,edx-ds,udemy-ds}. The ecosystem contains a wide range of tools such as \pandas for data manipulation and analysis, \matplotlib for data visualization, and \textsf{Scikit-learn} for machine learning, all aimed towards simplifying different stages of the data science pipeline. 

In this paper we focus on one part of the pipeline, exploratory data analysis (EDA) for statistical modeling, the process of understanding data through data manipulation and visualization. It is an essential step in every data science project\cite{Tessera21}. For statistical modeling, EDA often involves routine tasks such as understanding a single variable (univariate analysis),  understanding the relationship between two random variables (bivariate analysis), and understanding the impact of missing values (missing value analysis). 

 \add{Currently, there are two EDA solutions in Python. Each of them provide APIs in different granularity and have different drawbacks.}
 
 \begin{table}[t]
\small
\setlength\tabcolsep{2pt}
\footnotesize \caption{\add{Comparison of EDA solutions in Python.}}\vspace{-1em}
    \begin{tabular}{lccc}
    \hline
     & \textbf{\pandasviz} & \textbf{\pp} & \textbf{\edax}  \\ \hline
    \textbf{Easy to Use} & \large{$\times$} & \large{\checkmark} & \large{\checkmark}  \\ 
    \textbf{Interactive Speed} & \large{\checkmark} & \large{$\times$} & \large{\checkmark} \\ 
    \textbf{Easy to Customize} & \large{\checkmark} & \large{$\times$} & \large{\checkmark} \\ \hline
    \end{tabular}
    
    \label{tab:eda_comparison}\vspace{-3em}
\end{table}

\sloppy

\add{\emph{\pandasviz}. The first one is \pandasviz, where \pyviz represents a Python plotting library, such as \matplotlib~\cite{Hunter:2007}, \seaborn~\cite{waskom2020seaborn}, and \bokeh~\cite{bokeh}. Fundamentally, plotting libraries are \emph{not} designed for EDA but for plotting. Their APIs are at a very low level, hence they are not easy to use: To complete an EDA task, a data scientist needs to think about what plots to create, then using \pandas to manipulate the data so that it can be fed into a plotting library to create these plots. Often there is a gap between an EDA task and the available plots -- a data scientist must write lengthy and repetitive code to bridge the gap.} 

\begin{figure*}[!t]
    \centering
    \includegraphics[width=.88\textwidth]{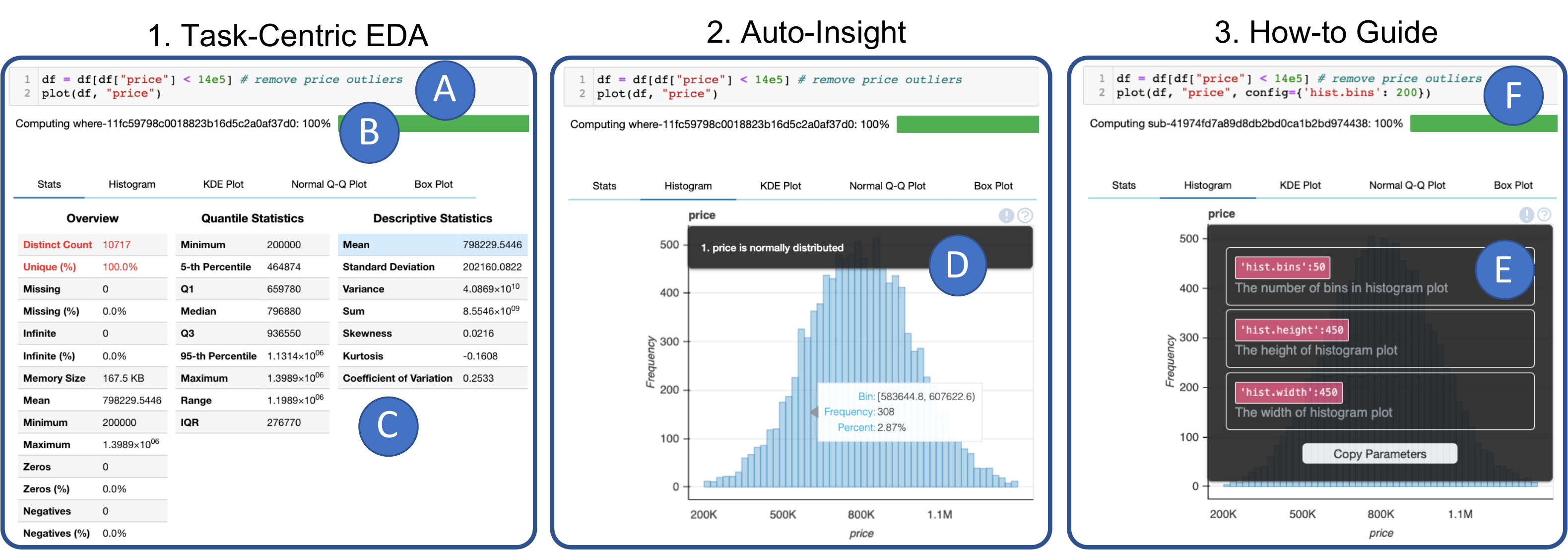}\vspace{-.5em}
    \vspace{-.5em}
    \caption{The front-end of \edax}
    \label{fig:user_exp}\vspace{-1em}
\end{figure*}

\fussy

\add{\emph{\pp}. The second one is \pp~\cite{pandasprofiling2019}. It provides a very high level API and allows a data scientist to generate a comprehensive profile report. The report has five main sections: \texttt{Overview}, \texttt{Variables}, \texttt{Interactions}, \texttt{Correlation}, and \texttt{Missing Values}. 
Its general utility makes it the most popular EDA library in Python. As of September, 2020, \pp had over 2.5M downloads on PyPI and over 5.9K GitHub stars.} 

\add{While \pp is effective for one-time profiling, it suffers from two limitations for EDA due to its high-level API design: (i) Firstly, it does not achieve interactive speed since generating a profile report often takes a long time. This is suffering as EDA is an \emph{iterative} process. Furthermore, the report shows information for all columns, potentially misdirecting the user and adding processing time. (ii) Secondly, it is not easy to customize a profile report. In a profile report, the plots are automatically generated thus it is very likely that the user wants to fine tune the parameters of each plot (e.g., the number of bins in a histogram). There could be hundreds of parameters associated with a profile report. It is not easy for users to figure out what they can customize and how to customize to meet their needs.}

\add{Table~\ref{tab:eda_comparison} summarizes the drawbacks of \pandasviz and \pp. The key challenge is how to overcome the limitations of existing tools and design a new EDA system that can achieve three design goals: easy to use, with interactive speed, and easy to customize. To address this challenge, we build \edax, a novel task-centric EDA system in Python. We identify a list of common EDA tasks for statistical modeling, mapping each task to a single function call through careful API design. As a result of this task-oriented approach, \edax affords many more \emph{fine-grained} tasks such as univariate analysis and correlation analysis. Figure~\ref{fig:user_exp}-1 illustrates how our example user might use \edax to do a univariate analysis task after removing their outliers. The analyst calls the \texttt{plot}(df, "price") in \edax, where df is a dataframe and "price" is the column name. \edax detects \texttt{price} as a numerical variable and automatically generates suitable statistics (e.g., max, mean, quantile) and plots (e.g., histogram, box plot), which help the user gain a deeper understanding of the \texttt{price} column quickly and effectively.}

\add{With the task-centric approach, \edax is able to achieve all three design goals: (i) \emph{Easy to Use.} Since each EDA task is directly mapped to a single function call, users only need to think about what tasks to work on rather than what to plot and how to plot. To further improve usability, we design an \emph{auto-insight} component to automatically highlight possible interesting patterns in visualizations. (ii) \emph{Interactive Speed.} Different from \pp, \edax supports fine-grained tasks thus it can avoid unnecessary computation on irrelevant information. To further improve the speed, we carefully design our data processing pipeline based on \dask, a scalable computing framework in Python. (iii) \emph{Easy to Customize.} With the task-centric API design, the parameters are grouped by different EDA tasks and each API only contains a small number of task-related parameters, making it much easier to customize. Besides, we implement a \emph{how-to guide} component in \edax to further improve the  customizability.}

We conduct extensive experiments to compare \edax with \pp. The performance results on 15 real-world datasets from Kaggle~\cite{kaggle} show that i) \edax responded to a fine-grained EDA task in seconds while \pp spent several orders of magnitude more time in creating a profile report on the same dataset; ii) if the task is to create a profile report, \edax was $4-20\times$ faster than \pp.  
Through a user study we show that i) real world participants of varying skill levels completed $2.05$ times more tasks on average with \edax than with \pp; ii) \edax helped participants answering $2.20$ times more correct answers.

The following summarizes our contributions: 
\vspace{-.5em}

\begin{itemize}[leftmargin=*]\itemsep0.35em
    \item We explore the limitations of existing EDA solutions in Python and propose a task-centric framework to overcome them.
    \item We design a task-centric EDA API for statistical modeling, allowing to declaratively specify an EDA task in one function call.
    \item We identify three challenges to implement \edax, and propose effective solutions to enhance the scalability, usability, and customizability of the system. 
    \item We conduct extensive experiments to compare \edax with \pp, the state-of-the-art EDA system in Python. The results show that \edax significantly outperforms \pp in speed, effectiveness, and user preference.  
\end{itemize}

\msection{Related Work}\label{sec:related-work}

\stitle{EDA Tools in Python and R.} Python and R are the two most popular programming languages in data science. Similar to Python, there are many EDA libraries in R including DataExplorer~\cite{dataexplorer} and visdat~\cite{tierney2017visdat} (see~\cite{staniak2019landscape} for a recent survey). However, they are either similar to \pandasviz or \pp, thus having the same limitations as them. 
In the database community, recently, there is a growing interest in building EDA systems for Python programmers in order to benefit a large number of real-world data scientists~\cite{deutch2020explained,lux}.  To the best our knowledge, \edax is the first task-centric EDA system in Python, and the only EDA system dedicated specifically to the notion of task-centric EDA.

\stitle{GUI-based EDA.} A GUI-based environment is commonly used for doing EDA, particularly among non-programmers. In such an environment, an EDA task is triggered by a click, drag, drop, etc (rather than a Python function call). Many commercial systems including Tableau~\cite{tableau}, Excel~\cite{excel}, Spotfire~\cite{spotfire}, Qlik~\cite{qlik}, Splunk~\cite{splunk}, Alteryx~\cite{alteryx}, \add{SAS~\cite{sas}, JMP~\cite{jmp} and SPSS~\cite{spss} support doing EDA using a GUI. Although these systems are suitable in many cases, they all have the fundamental limitations of being removed from the programming environment and lacking flexibility.}

In recent years, there has been abundant research in visualization recommendation systems~\cite{wongsuphasawat2015voyager, wongsuphasawat2017voyager,mackinlay2007show,cui2019datasite,siddiqui2016zenvisage, dibia2019data2vis,hu2019vizml,luo2018deepeye,2019-draco}. Visualization recommendation is the process of automatically determining an interesting visualization and presenting it to the user. Another related area is automated insight generation (auto-insights). An auto-insight system mines a dataset for statistical properties of interest~\cite{ding2019quickinsights, lin2018bigin4, tang2017topkinsights,powerbi,demiralp2017foresight,vartak2015seedb}.  Unlike these systems, \edax is a programming-based EDA tool that has several advantages over GUI-based EDA systems including seamless integration in the Python data science ecosystem, and flexibility since the data scientist is not restricted to one GUI's functionalities. 

\stitle{Data Profiling.} Data profiling is the process of deriving summary information (e.g., data types, the number of unique values in columns) from a dataset (see~\cite{abedjan2015profiling} for a data profiling survey). Metanome~\cite{papenbrock2015data} is a data profiling platform where the user can run profiling algorithms on their data to generate different summary information. Data profiling can be used in the tasks of data quality assessment (e.g., Profiler~\cite{kandel2012profiler}) and data cleaning (e.g., Potter's Wheel~\cite{raman2001potter}). Although \edax performs data profiling, unlike the above systems it is integrated effectively in a Python programming environment.

Python data profiling tools including Pandas-profiling~\cite{pandasprofiling2019}, Sweetviz~\cite{sweetviz}, and AutoViz~\cite{autoviz}, enable profiling a dataset by running one line of code. These systems provide rigid and coarse-grained analysis which are not suitable for general, ad-hoc EDA.
\msection{Task-Centric EDA}
\label{sec:task-centric}

In this section, we first introduce common EDA tasks for statistical modeling, and then describe our task-centric EDA API design.

\begin{figure*}[!t]
    \centering
    \includegraphics[width=.98\textwidth]{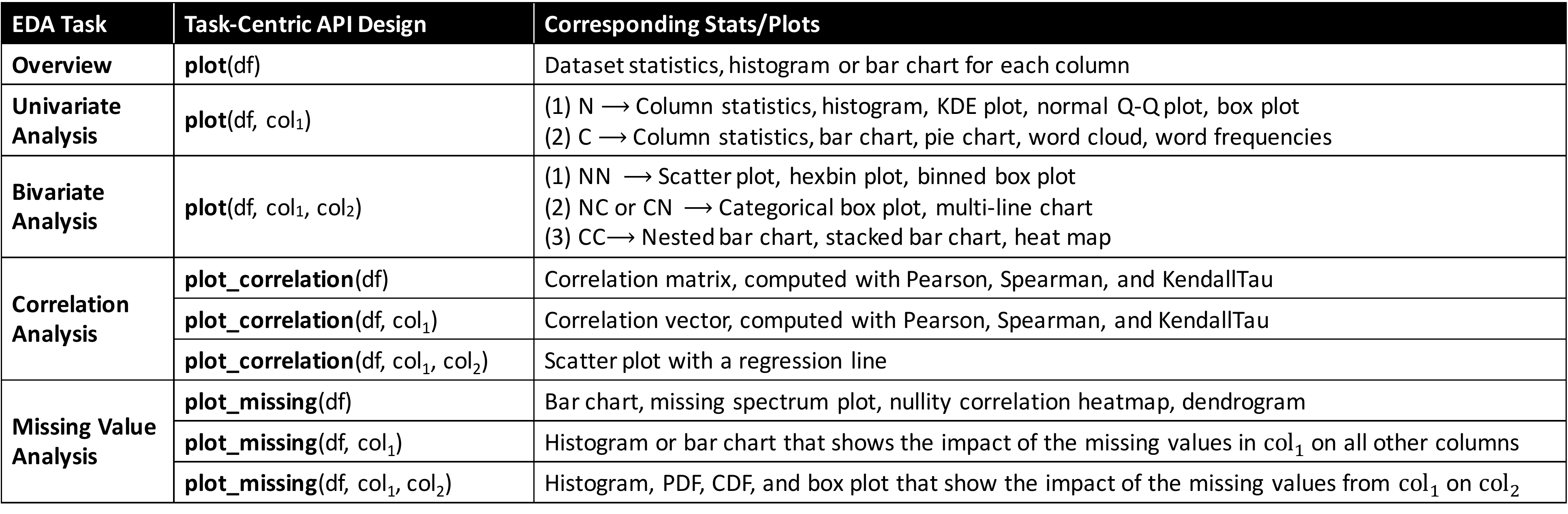}
     \vspace{-1em}
    \caption{A set of mapping rules between EDA tasks and corresponding stats/plots (N = Numerical, C = Categorical)}
    \label{fig:mapping}\vspace{-1em}
\end{figure*}

\msubsection{{\large Common EDA Tasks for Statistical Modeling}}

\add{Inspired by the profile report generated by \pp and existing work~\cite{peng2012exploratory, seltman2012experimental, bilogur2018missingno}, we identify five common EDA tasks}. We will use a running example to illustrate why they are needed in the process of statistical modeling.

Suppose a data scientist wants to build a regression model to predict house prices. The training data consists of four features (\texttt{size}, \texttt{year\_built}, \texttt{city}, and \texttt{house\_type}) and the target (\texttt{price}).

\begin{itemize}[leftmargin=*]\itemsep0.35em
    \item \textbf{Overview}. At the beginning, the data scientist has no idea about what's inside the dataset, so she wants to get a quick overview of the entire dataset. This involves computing some basic statistics and creating some simple visualizations. For example, she may want to check the number of features, the data type of each feature (numerical or categorical), and create a histogram for each numerical feature and a bar chart for each categorical feature. 
    \item \textbf{Correlation Analysis}. To select important features or identify redundant features, correlation analysis is commonly used. It computes a correlation matrix, where each cell in the matrix represents the correlation between two columns. A correlation matrix can show which features are highly correlated with the target and which two features are highly correlated with each other. For example, if the feature, \texttt{size}, is highly correlated with the target, \texttt{price}, then knowing \texttt{size} will reveal a lot of information about \texttt{price}, thus it is an important feature. If two features, \texttt{city} and \texttt{house\_type}, are highly correlated, then one of the features is redundant and can be removed
    \item \textbf{Missing Value Analysis}. It is more common than not for a dataset to have missing values. The data scientist needs to create customized visualizations to understand missing values. For example, she may create a bar chart, which depicts the amount of missing values in each column, or a missing spectrum plot, which visualizes which rows has more missing values. 
    \item \textbf{Univariate Analysis}. Univariate analysis aims to gain a deeper understanding of a single column. It creates various statistics and visualizations of that column. For example, to deeply understand the feature \texttt{year\_built}, the data scientist may want to compute the min, max, distinct count, median, variance of \texttt{year\_built}, and create a box plot to examine outliers, a normal Q-Q plot to compare its distribution with the normal distribution.
    \item \textbf{Bivariate Analysis}. Bivariate analysis is to understand the relationship between two columns (e.g., a feature and the target). There are many visualizations to facilitate the understanding. For example, to understand the relationship between \texttt{year\_built} and \texttt{price}, she may want to create a scatter plot to check whether they have a linear relationship, and a hexbin plot to check the distribution of \texttt{price} in different year ranges. 
\end{itemize}

\add{There are certainly other EDA tasks used for statistical modeling, however we have opted to focus on the main tasks systems such as \pp commonly present in their reports. This allows us to make a fair comparison between our system design approaches. In the future we intend to address more tasks, such as time-series analysis and multi-variate analysis (more than two variables).}

\msubsection{\edax's Task-Centric API Design}

\add{The goal of our API design is to enable the user to trigger an EDA task through a single function call. We consider simplicity and consistency as the principle of API design. The simple and consistent API makes our system more accessible in practice~\cite{buitinck2013api}. However, it is challenging to design simple and consistent APIs for a variety of EDA tasks. Our key observation is that the EDA tasks for statistical modeling tend to follow a similar pattern~\cite{wickham2016r}: start with an overview analysis and then dive into detailed analysis. Hence, we design the API in the following form: }
\add{$$\textbf{plot}\_\textbf{\emph{tasktype}}(\textsf{df}, \textsf{col\_list}, \textsf{config}),$$}%
\add{where \textsf{plot\_\textsf{\emph{tasktype}}} is the function name,  \textsf{tasktype} is a concise description of the task, the first argument is a DataFrame, the second argument is a list of column names, and the third argument is a dictionary of configuration parameters. If column names are not specified, the task will be performed on all the columns in the DataFrame (overview analysis); otherwise, it will be performed on the specified column(s) (detailed analysis). This design makes the API extensible, i.e., it is easy to add an API for a new task.}

Following this pattern, we design three functions in \edax to support the five EDA tasks:

\stitle{plot.} We use the \texttt{plot}$(\cdot)$ function with different arguments to represent the overview task, the univariate analysis task, and the bivariate analysis task, respectively. To understand how to perform EDA effectively with this function, the following gives the syntax of the function call with the intent of the data scientist:
\begin{itemize}
    \item plot(df): ``I want an overview of the dataset''
    \item plot(df, \texttt{col}$_1$): ``I want to understand \texttt{col}$_1$''
    \item plot(df, \texttt{col}$_1$, \texttt{col}$_2$): ``I want to understand the relationship between \texttt{col}$_1$ and \texttt{col}$_2$''
\end{itemize}

\stitle{plot\_correlation.} The \texttt{plot\_correlation}$(\cdot)$ function triggers the correlation analysis task. The user can get more detailed correlation analysis results by calling \texttt{plot\_correlation}(df, col$_1$) or \texttt{plot\_correlation}(df, col$_1$, col$_2$). 
\begin{itemize}
    \item plot\_correlation(df): ``I want an overview of the correlation analysis result of the dataset''
    \item plot\_correlation(df, \texttt{col}$_1$): ``I want to understand the correlation between \texttt{col}$_1$ and the other columns''
    \item plot\_correlation(df, \texttt{col}$_1$, \texttt{col}$_2$): ``I want to understand the correlation between \texttt{col}$_1$ and \texttt{col}$_2$''
    \vspace{-.25em}
\end{itemize}

\stitle{plot\_missing.} The \texttt{plot\_missing}$(\cdot)$ function triggers the missing value analysis task. Similar to \texttt{plot\_correlation}$(\cdot)$, the user can call \texttt{plot\_missing}(df, col$_1$) or \texttt{plot\_missing}(df, col$_1$, col$_2$) to get more detailed analysis results. 
\begin{itemize}
    \vspace{-.25em}
    \item plot\_missing(df): ``I want an overview of the missing value analysis result of the dataset''
    \item plot\_missing(df, \texttt{col}$_1$): ``I want to understand the impact of removing the missing values from \texttt{col}$_1$ on other columns''
    \item plot\_missing(df, \texttt{col}$_1$, \texttt{col}$_2$): ``I want to understand the impact of removing the missing values from \texttt{col}$_1$ on \texttt{col}$_2$''
    \vspace{-.25em}
\end{itemize}

The key observation that makes task-centric EDA possible is that there are nearly universal kinds of stats or plots that analysts employ in a given EDA task. For example, if the user wants to perform univariate analysis on a numerical column, she will create a histogram to check the distribution, a box plot to check the outliers, a normal Q-Q plot to compare with the normal distribution, etc. Based on this observation, we pre-define a set of mapping rules as shown in Figure~\ref{fig:mapping}, where each rule defines what stats/plots to create for each EDA task. Once a function, e.g., \texttt{plot(df,"price")}, is called, \edax first detects the data type of \texttt{price}, which is numerical. Based on the second row in Figure~\ref{fig:mapping}, since col$_1$ = N, \edax will automatically generate the column statistics, histogram, KDE plot, normal Q-Q plot, and box plot of \texttt{price}.

\add{The mapping rules are selected from existing literature and open-source tools in the statistics and machine learning community. For univariate, bivariate, and correlation analysis, we refer to the data-to-viz project~\cite{data-to-viz}, ‘Exploratory Graphs’ section in~\cite{peng2012exploratory} and Section 4 in~\cite{seltman2012experimental}. For overview analysis and missing value analysis, the mapping rules are derived from the \pp library and the Missingno library~\cite{bilogur2018missingno}, respectively. \edax also leverages the open-source community to keep adding and improving its rules. For example, one user has created an issue in our GitHub repository to suggest adding violin plots to the \texttt{plot(df,x)} function. As \edax is being used by more users, we expect to see more suggestions like this in the future.}

\msection{System Architecture}\label{sec:architecture}

\begin{figure*}[!t]
    \centering
    \includegraphics[width=1\textwidth]{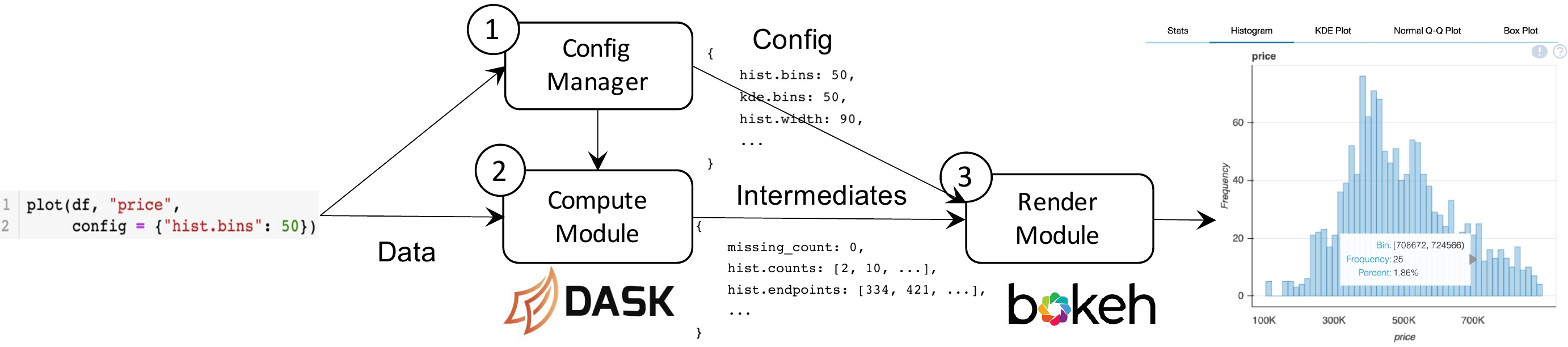}\vspace{-1em}
    \caption{The \edax system architecture}\vspace{-1em}
    \label{fig:sys_arch}
\end{figure*}

This section describes \edax's front-end user experience and introduces the back-end system architecture. 

\msubsection{Front-end User Experience}

To demonstrate \edax, we will continue our house price prediction example, using \edax to assist in removing outliers from the \texttt{price} variable, assessing the resulting distribution, and determining how to further customize the analysis.  Figure~\ref{fig:user_exp} depicts the steps to perform this task using \pandas and \edax in a Jupyter notebook. Part~\circleddark{A} shows the required code: in line~1, the records with an outlying price value are removed (the threshold is \$1,400,000), and in line 2 the \edax function \texttt{plot} is called to analyze the filtered distribution of the variable \texttt{price}. Part~\circleddark{B} shows the progress bar. Part~\circleddark{C} shows the default output tab which consists of tables containing various statistics of the column's distribution. Note that each data visualization is output in a separate panel, and tabs are used to navigate between panels.

\begin{itemize}[leftmargin=*]\itemsep0.35em
    \item  \textbf{Auto-Insight}. If an insight is discovered by \edax, a (\circleddark{!}) icon will be shown on the top right corner of the associated plot. Part~\circleddark{D} shows an insight associated with the histogram plot: \texttt{price} is normally distributed.
    \item  \textbf{How-to Guide}. A how-to guide will pop up after clicking a (\circleddark{?}) icon. As shown in Part~\circleddark{E}, it contains the information about customizing the associated plot. In this example, the data scientist may want to create a histogram with more bins, so she can copy the  code (\texttt{"hist.bins":50}) in the how-to guide, paste it as a parameter to the plot function, and increase the number of bins from 50 to 200 as shown in Part~\circleddark{F}.
\end{itemize}

\msubsection{Back-end System Architecture} \label{arch:sys_comps}

The \edax back-end is presented in Figure \ref{fig:sys_arch}, consisting of three components: \circled{1} The Config Manager configures the system's parameters, \circled{2} the Compute module performs the computations on the data, and \circled{3} the Render module creates the visualizations and layouts. The Config Manager is used to organize the user-defined parameters and set default parameters in order to avoid setting and passing many parameters through the Compute and Render modules. The separation of the Compute module and the Render module has two benefits: First, the computations can be distributed to multiple visualizations. For example, in \texttt{plot(df, col$_1$=N)} in Figure~\ref{fig:mapping}, the column statistics, normal Q-Q plot, and box plot all require quantiles of the distribution. Therefore, the quantiles are computed once and distributed appropriately to each visualization. Second, the intermediate computations (see Section~\ref{arch:compute}) can be exposed to the user. This allows the user to create the visualizations with her desired plotting library.

\vspace{-.5em}
\subsubsection{Config Manager}
The Config Manager (\circled{1} in Figure~\ref{fig:sys_arch}) sets values for all configurable parameters in \edax, and stores them in a data structure, called the \texttt{config}, which is passed through the rest of the system. Many components of \edax are configurable including which visualizations to produce, the insight thresholds (see Section~\ref{arch:compute}), and visualization customizations such as the size of the figures. In Figure \ref{fig:sys_arch}, the \texttt{plot} function is called with the user specification \texttt{bins=50}; the Config Manager sets each bin parameter to have a value of 50, and default values are set for parameters not specified by the user. The \texttt{config} is then passed to the Compute and Render modules and referenced when needed.

\sloppy
\vspace{-.5em}
\subsubsection{Compute module} \label{arch:compute}
The Compute module takes the data and \texttt{config} as input, and computes the \texttt{intermediates}. The \texttt{intermediates} are the results of all the computations on the data that are required to generate the visualizations for the EDA task. Figure~\ref{fig:sys_arch} shows example \texttt{intermediates}. The first element is the count of missing values which is shown in the \texttt{Stats} tab, and the next two elements are the counts and bin endpoints of the histogram. Such statistics are ready to be fed into a visualization.

\fussy

Insights are calculated in the Compute module. A data fact is classified as an insight if its value is above a threshold (each insight has its own, user-definable threshold). For example, in Figure~\ref{fig:user_exp} (Part \circled{B}), the distinct value count is high, so the entry in the table is highlighted red to alert the user about this insight. \edax supports a variety of insights including data quality insights (e.g., missing, infinite values), distribution shape insights (e.g., uniformity, skewness) and whether two distributions are similar.

We developed two optimization techniques to increase performance. First, we share computations between multiple visualizations as described in the beginning of Section~\ref{arch:sys_comps}. Second, we leverage Dask to parallelize computations (see Section~\ref{sec:implementation} for details).

\vspace{-.5em}
\subsubsection{Render module}

The last system component is the Render module, which converts the \texttt{intermediates} into data visualizations. There is a plethora of Python plotting libraries (e.g., \matplotlib, \seaborn, and \bokeh), however, they provide limited or no support for customizing a plot's \emph{layout}. A \textit{layout} is the surrounding environment in which visualizations are organized and embedded. Our layouts need to consolidate many elements including charts, tables, insights, and how-to guides. To meet our needs, we use the library \bokeh to create the plots, and embed them in our own HTML/JS layout.

\msection{Implementation}\label{sec:implementation}
In this section, we introduce the detailed implementation of \edax's Compute module. We first introduce the background of \dask and discuss why we choose \dask as the back-end engine. We then present our ideas for using \dask to optimize \edax.

\msubsection{Why \dask}
\label{sec:dask-bkg}

\stitle{\dask Background.}
\dask is an open source library providing scalable analytics in Python. It offers similar APIs and data structures with other popular Python libraries, such as \numpy, \pandas, and \scikitlearn. Internally, it partitions data into chunks, and runs computations over chunks in parallel. 

The computations in \dask are lazy. \dask will first construct a computational graph that expresses the relationship between tasks. Then, it optimizes the graph to reduce computations such as removing unnecessary operators. Finally, it executes the graph when an eager operation like \texttt{compute} is called.

\stitle{Choice of Back-end Engine.} \add{ We use \dask as the back-end engine of \edax for three reasons: (i) it is lightweight and fast in a single-node environment, (ii) it can scale to a distributed cluster, and (iii) it can optimize the computations required for multiple visualizations via lazy evaluation. We considered other engines like Spark variants~\cite{DBLP:conf/hotcloud/ZahariaCFSS10, koalas} (PySpark and Koalas) and Modin~\cite{DBLP:journals/corr/abs-2001-00888}, but found that they were less suitable for \edax than \dask. Since Spark is designed for computations on very big data (TB to PB) in a large cluster, PySpark and Koalas are not lightweight like \dask and have a high scheduling overhead on a single node. For Modin, most of its operations are eager, so for each operation a separate computational graph is created. This approach does not optimize across operations, unlike \dask's approach. In Section~\ref{sec:exp-more-eval}, we further justify our choice to use \dask experimentally.}

\msubsection{Performance Optimization}

Given an EDA task, e.g., \texttt{plot\_missing(df)}, we discuss how to efficiently compute its \texttt{intermediates} using \dask. We observe that there are many redundant computations between visualizations. For example, as shown in Figure~\ref{fig:mapping}, \texttt{plot\_missing(df)} creates four visualizations (bar chart, missing spectrum plot, nullity correlation heatmap, and dendrogram). They share many computations, such as computing the number of rows, checking whether a cell is missing or not. To leverage \dask to remove redundant computations, we seek to express all the computations in a \emph{single} computational graph.  To implement this idea, we can make all the computations lazy and call an eager operation at the end. In this way, \dask will optimize the whole graph before actual computations happen. 

However, there are several issues with this implementation. In the following, we will discuss them and propose our solutions.

\stitle{Dask graph fails to build.} The first issue is that the \texttt{rechunk} function in \dask cannot be directly incorporated into the big computational graph. This is because that its first argument, a \dask array, requires knowing the chunk size information, i.e., the size of each chunk in each dimension. If \texttt{rechunk} was put into our computational graph, an error would be raised since the chunk size information is unknown for a delayed \dask array. 

Since \texttt{rechunk} is needed in multiple \texttt{plot} functions in \edax, we have to address this issue. One solution is to replace the \texttt{rechunk} function call in each \texttt{plot} function with the code written by the low-level \dask task graph API. However, this solution has a high engineering cost, which requires writing hundreds of lines of \dask code. It also has a high maintenance cost compared to using the \dask built-in \texttt{rechunk} function. 

We propose to add an additional stage before constructing the computational graph. In this stage, we precompute the chunk size information of the dataset and pass the precomputed chunk size to the \dask graph. In this way, the \dask graph can be constructed successfully by adding one line of code.

\stitle{\dask is slow on tiny data.} Although putting all possible operations in the graph can fully leverage \dask's optimizations, it also increases the overhead caused by scheduling. When data is large, the scheduling overhead is negligible compared to the computing overhead. However, when data is tiny, the scheduling may be the bottleneck and using \dask is less efficient than using \pandas. 

For the nine \texttt{plot} functions in Figure~\ref{fig:mapping}, we observe that they all follow the same pattern: the computational graph takes as input a DataFrame (large data) and continuously reduces its size by aggregation, filtering, etc. Based on this observation, we separate the computational graph into two parts: \emph{\dask Computation} and \emph{\pandas Computation}. In the \emph{\dask Computation}, the data is computed in \dask and the result is transformed into a \pandas DataFrame. In the \emph{\pandas Computation}, it takes the DataFrame as input and does some further processing to generate the intermediate results, which will be used to create visualizations. 

\final{Currently, we heuristically determine the boundary between the two parts. For example, the computation for plot\_correlation(df) is separated into two stages. In the first stage we use \dask to compute the correlation matrix from the user input and then in the second stage we use \pandas to transform and filter the correlation matrix. This is because for a dataset with $n$ rows and $m$ columns, it is usually the case that $n >> m$. As a result, it would be beneficial to let \pandas handle the correlation matrix, which has the size $m \times m$. } Since we only need to handle nine \texttt{plot} functions, it is still manageable. We will investigate how to automatically separate the two parts in the future.

\begin{figure}[!t]
    \centering
    \includegraphics[width = 0.5\textwidth]{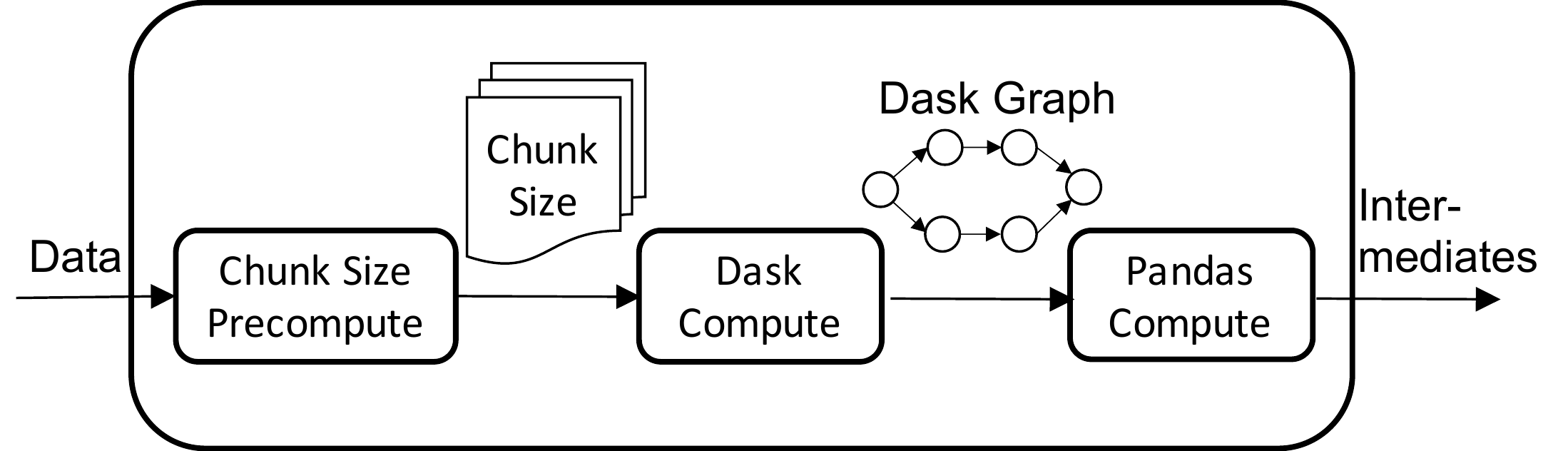}
    \vspace{-1.5em}
    \caption{Data processing pipeline in the Compute module}
    \label{fig:compute}
    \vspace{-.5em}
\end{figure}

\sloppy 

\fussy

\stitle{Putting everything together.} Figure~\ref{fig:compute} shows the data processing pipeline in the Compute module of \edax. When data comes, \edax first precomputes the chunk size information using \dask. After that, it constructs the computational graph using \dask again with the precomputed chunk size filled. Then, it computes the graph. After that, it transforms the computed data into \pandas and finishes the \pandas computation. In the end, the \texttt{intermediates} are returned.

\msection{Experimental Evaluation}\label{sec:experiment}
In this section, we conduct extensive experiments to evaluate the efficiency and the user experience of \edax.

\msubsection{Performance Evaluation}
We first evaluate the performance of \edax by comparing its report functionality with \pp. Afterwards, we conduct a self comparison, which evaluates each plot function with different variations, in order to gain a deep understanding of the performance. We used 15 different datasets varying in number of rows, number of columns and categorical/numerical column ratio, as listed in \Cref{tab:datasets}. The experiments were performed on an Ubuntu 16.04 Linux server with 64 GB memory and 8 Intel E7-4830 cores.

\stitle{Comparing with \pp.} To test the general efficiency of \edax, we compared the end-to-end running time of generating reports over the 15 datasets in both tools. For \pp, we first used \texttt{read\_csv} from Pandas to read the data, then we created a report in \pp with PhiK, Recoded and Cramer's V correlations disabled (since \edax does not implement these correlation types).
For \edax, we used \texttt{read\_csv} from \dask to read the dataset. Since it is a one-time task to create a profiling report, loading the data using the most suitable function for each tool is reasonable. Results are shown in \Cref{tab:datasets}. We observe that using \edax to generate reports is 4x $\sim$ 20x faster compared to \pp in general. The acceleration mainly comes from the optimization to make the tasks into a single \dask graph so that they can be fully parallelized. We also observe that \edax usually gains more performance compared to \pp on numerical data and data with fewer categorical columns (driving performance on credit, basketball, and diabetes).

\begin{table}[!t]
\small
    \setlength\tabcolsep{1.45pt}
    \caption{Comparing \edax with \pp on 15 real-world data science datasets from Kaggle (\textbf{N} = Numerical, \textbf{C} = Categorical, PP=\pp).}\vspace{-1em}
    \begin{tabular}{lllllcl}
    \hline
    \textbf{\small Dataset} & \textbf{\small Size} & \textbf{\small \#Rows} & \textbf{\small \#Cols (N/C)} & \textbf{\small PP} & \textbf{\small DataPrep} & \textbf{Faster} \\ \hline
    heart~\cite{dat:heart}            & 11KB          & 303             & 14 (14/0)       & 17.7s                      & 2.0s           & 8.6$\times$               \\
    diabetes~\cite{dat:diabetes}         & 23KB          & 768             & 9 (9/0)         & 28.3s                      & 1.6s           & 17.7$\times$              \\
    automobile~\cite{dat:automobile}       & 26KB          & 205             & 26 (10/16)      & 38.2s                      & 3.9s           & 9.8$\times$              \\
    titanic~\cite{dat:titanic}          & 64KB          & 891             & 12 (7/5)        & 17.8s                      & 2.1s           & 8.5$\times$              \\
    women~\cite{dat:women}            & 500KB         & 8553            & 10 (5/5)        & 19.8s                      & 2.3s           & 8.6$\times$              \\
    credit~\cite{dat:credit}           & 2.7MB         & 30K             & 25 (25/0)       & 127.0s                     & 6.1s           & 20.8$\times$             \\
    solar~\cite{dat:solar}            & 2.8MB         & 33K             & 11 (7/4)        & 25.1s                      & 2.7s           & 9.3$\times$              \\
    suicide~\cite{dat:suicide}          & 2.8MB         & 28K             & 12 (6/6)        & 20.6s                      & 2.8s           & 7.4$\times$              \\
    diamonds~\cite{dat:diamonds}         & 3MB           & 54K             & 11 (8/3)        & 28.2s                      & 3.1s           & 9$\times$                \\
    chess~\cite{dat:chess}            & 7.3MB         & 20K             & 16 (6/10)       & 23.6s                      & 4.3s           & 5.5$\times$              \\
    adult~\cite{dat:adult}            & 5.7MB         & 49K             & 15 (6/9)        & 23.2s                      & 4.0s           & 5.8$\times$              \\
    basketball~\cite{dat:basketball}       & 9.2MB         & 53K             & 31 (21/10)      & 126.2s                     & 9.9s           & 12.7$\times$             \\
    conflicts~\cite{dat:conflicts}        & 13MB          & 34K             & 25 (10/15)      & 34.9s                      & 8.6s           & 4$\times$                \\
    rain~\cite{dat:rain}             & 13.5MB        & 142K            & 24 (17/7)       & 100.1s                     & 11.6s          & 8.6$\times$              \\
    hotel~\cite{dat:hotel}            & 16MB          & 119K            & 32 (20/12)      & 83.2s                      & 13s            & 6.4$\times$  \\\hline
    \end{tabular}
    
    \label{tab:datasets}\vspace{-1.5em}
\end{table}

\sloppy

\begin{figure}[!tbp]
    \centering
    \includegraphics[width=0.5\textwidth]{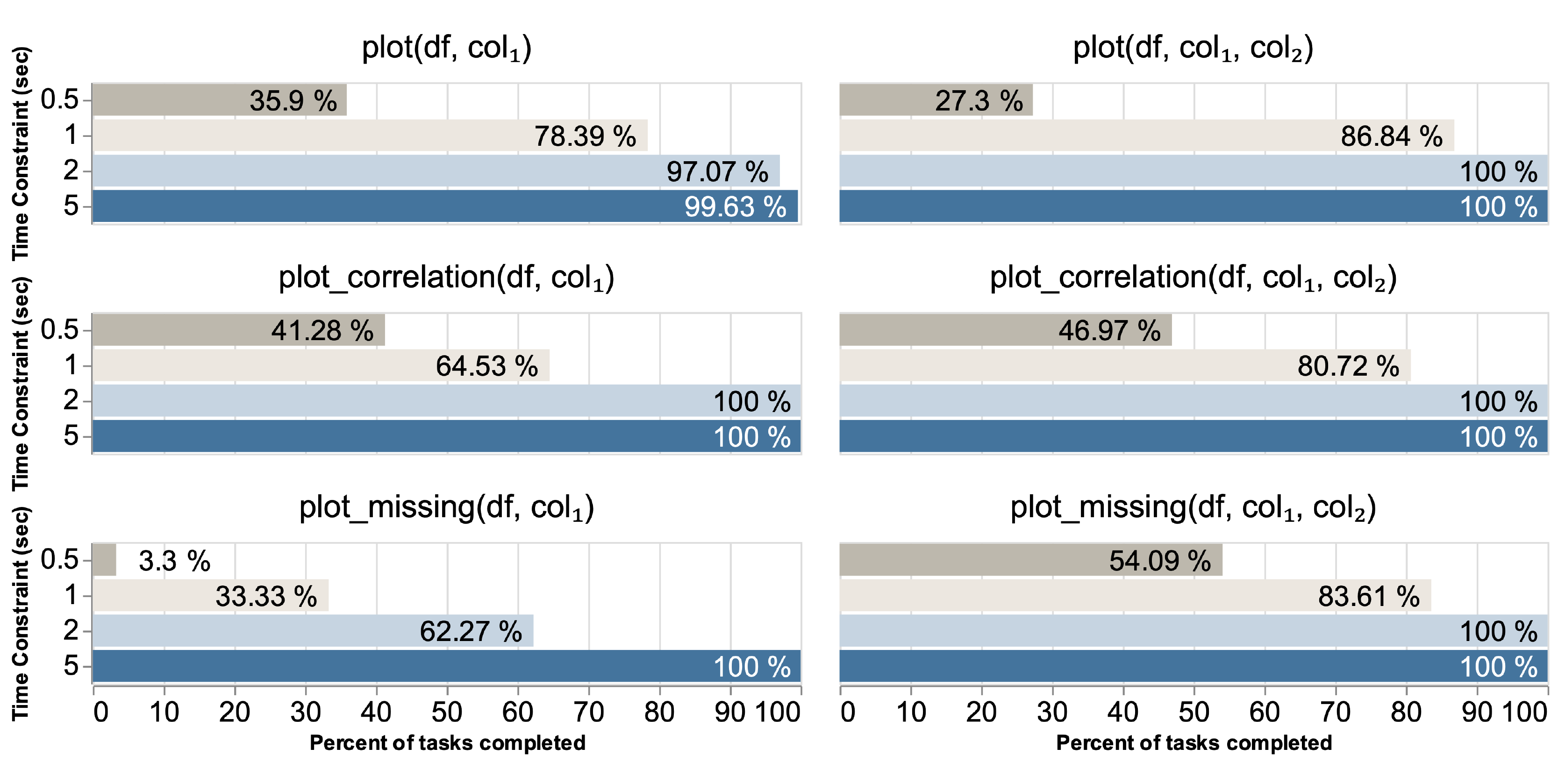}\vspace{-1em}
    \caption{The percentage of tasks to finish within the given time constraint.}
    \vspace{-1em}
    \label{fig:self_comp_drilldown}
\end{figure}

\stitle{Self-comparison.} We analyzed the running time of \edax's functions to determine whether they can complete within a reasonable response time for interactivity. We ran \texttt{plot()}, \texttt{plot\_correlation()}, and \texttt{plot\_missing()} for each column in each dataset, and we ran the three functions for all unique pairs of columns in each dataset (limited to categorical columns with no more than 100 unique values for \texttt{plot(df, col$_1$, col$_2$)} so the resulting visualizations contain a reasonable amount of information). Figure \ref{fig:self_comp_drilldown} shows the percent of total tasks for each function that finish within 0.5, 1, 2 and 5 seconds. Note that dataset loading time is also included in the reported run times. The majority of tasks completed within 1 second for each function except \texttt{plot\_missing(df, col$_1$)}. \texttt{plot\_missing(df, col$_1$)} is computationally intensive because it computes two frequency distributions for each column (before and after dropping the missing values in column col$_1$).

\msubsection{Experiments on Large Data}
\label{sec:exp-more-eval}

\add{We conduct experiments to justify the choice of \dask (see Section~\ref{sec:dask-bkg}) and evaluate the scalability of \edax. We use the bitcoin dataset~\cite{dat:bitcoin} which contains 4.7 million rows and 8 columns.}

\begin{figure*}[!t]
    \centering
    \includegraphics[width=1\textwidth]{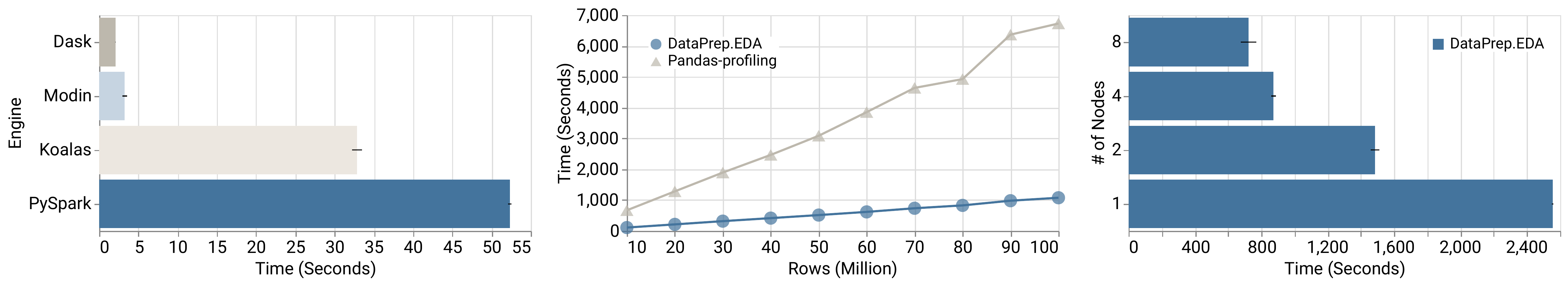}\vspace{-.5em}
    \hspace*{0em} (a) Comparing Engines (Single Node) \hspace*{6em}    (b) Varying Data Size (Single Node)   \hspace*{6em}   (c) Varying \# of Nodes \vspace{-1em}
    \caption{\add{Experiments on the Bitcoin Dataset: (a) Comparing the running time of using different engines to compute visualizations in plot(df); (b) Comparing the running time of create\_report(df) of \edax and \pp by varying data size; (c) Evaluating the running time of create\_report(df) of \edax by varying the number of nodes.}}\vspace{-1em}
    \label{fig:more_eval}
\end{figure*}

\stitle{Comparing Engines.} \add{We compare the time required for \dask, Modin, Koalas, and PySpark to compute the \texttt{intermediates} of \texttt{plot}(df). The results are shown in Figure~\ref{fig:more_eval}(a). The reason why \dask is the fastest is explained in Section~\ref{sec:dask-bkg}: Modin eagerly evaluates the computations and does not make full use of parallelization when computing multiple visualizations, and Koalas/PySpark have a high scheduling overhead in a single-node environment.}

\stitle{Varying Data Size.} \add{To evaluate the scalability of \edax, we compare the report functionality of \edax with \pp and vary the data size from 10 million to 100 million rows. The data size is increased by repeated duplication.
The results are shown in Figure~\ref{fig:more_eval}(b). Both \edax and \pp scale linearly, but \edax is around six times faster. 
This is because \edax leverages lazy evaluation to express all the computations in a single computational graph so that the computations can be fully parallelized by \dask.
}

\stitle{Varying \# of Nodes.} \add{To evaluate the performance of \edax in a cluster environment, we run the report functionality on a cluster and vary the number of nodes. The cluster consists of a maximum of 8 nodes, each with 64GB of memory and 16 2.0GHz E7-4830 CPUs dedicated to the \dask workers. There are also HDFS services running on the 8 nodes for data storage; the memory for these services is not shared with the \dask workers. The data is fixed at 100 million rows and stored in HDFS. We do not compare with \pp since it cannot run on a cluster. The result is shown in Figure~\ref{fig:more_eval}(c). We can see that \edax is able to run on a cluster and achieves better performance as increasing the number of nodes. This is because adding more compute nodes can reduce the I/O cost of reading data from HDFS. It is worth noting that the 1 worker setting in Figure~\ref{fig:more_eval}(c) is different from the single node setting in Figure~\ref{fig:more_eval}(b) where the former needs to read data from HDFS while the latter reads data from a local disk. Therefore, the 1 worker setting took longer to process 100 million rows than the single node setting.}

\msubsection{User Study}

Finally, we conducted a user study to validate the usability of our tool and its utility in supporting analytics. We focus on two questions using \pp as a comparison baseline: (1)~For different groups of participants, how do they benefit from the task-centric features introduced by \edax, and (2)~Does \edax reduce participants' false discovery rate. We hypothesize that the intuitive API in \edax will lead participants to complete more tasks in less time versus \pp, and that the additional tools provided by \edax will help participants to reduce their false discovery rate. 

\vspace{.25em}

\stitle{Methodology.}~\add{In our study, all recruited participants used both \edax and \pp to complete two different sets of tasks in a 50 minute session with software logging: one tool is used for one set of tasks.} Afterwards, they assessed both systems in surveys. Our study employs two datasets paired with task sets: (1) BirdStrike: $12$ columns related to bird strike damage on airplanes. The dataset compiles approximately $220,000$ strike reports from $2,050$ USA airports and $310$ foreign airports. (2) DelayedFlights: $14$ columns related to the causes of flight cancellations or delays. The dataset is curated by the Department of Transportation of United States, and contains $5,819,079$ records of cancelled flights. 

We followed a within-subjects design, with all participants making use of either tool to complete one set of tasks. We counterbalanced our allocation to account for all tool-dataset permutations and order effects. Within each task, participants finished five sequential tasks using one tool. As experience might influence how much an individual benefits from each tool, we recruited both skilled and novice analysts using a pre-screen concerning knowledge of python, data analysis, and the datasets in the study. To make sure that all participants had base knowledge of how to use both \pp and \edax, we gave participants with two introductory videos, a cheat sheet, and API documentation. 

\add{Participants were asked to complete 5 tasks sequentially using the data analysis tool. The tasks are designed to evaluate different functions provided by \edax, which is similar to the design of existing work~\cite{battle2019characterizing}.} They cover a relatively wide spectrum of the kinds of tasks that are frequently encountered in EDA, including gathering descriptive multivariate statistics of one or multiple columns, identifying missing values, and finding correlations. Though datasets have their own specific task instructions, each of the respective items shares the same goal across both datasets. For example, the first task of both sessions asks participants to investigate data distribution over multiple columns. 

In Task 1-3, participants use the provided tool to analyze the distribution of single or multiple columns. Participants conduct a univariate analysis in task 1 and a variate analysis in task 2. The task 3 asked the participant to examine distribution skewness.  
Task 4 examines missing values and their impact. Participants are expected to examine the distribution of missing values to come to a conclusion. Task 5 asks users to find columns with high correlation.

We use the fraction $\frac{\# correct answers}{\# completed tasks}$ to show the \emph{relative accuracy} of participants, since we noticed that a number of participants failed to finish tasks. 
Compared to traditional accuracy, relative accuracy therefore may better demonstrate positive discovery rate~\cite{card1997structure}.

\stitle{Examining Participants' Performance.}~ We examine accuracy and completed tasks to evaluate system performance.  The average number of completed tasks per participant using \edax (M:4.02, SD:1.21) was \emph{\textbf{2.05} } times higher than that using \pp (M:1.96, SD: 2.59, t(29)=5.26, $\rho<.00001$). When we compared skill levels, we could detect no difference (t(14)=$.882998$, $\rho<.441499$). Together, this suggests that \edax generally improved participants' efficiency in performing EDA tasks. 
Factoring in dataset, we find that \pp performs better in small datasets(M:3, SD:4.13) compared to a more complex one (M:1.1, SD:1.20, t(14)=$-3.26062$, $\rho<.0028$). As dataset complexity grows, \pp fails to scale up. No participants finished all tasks, and 42\% finished at most one task using \pp. On the other hand, 35\% of \edax participants finished all five tasks for the \emph{delayed} dataset. We did not observe a dataset difference for \edax(t(14)=$-0.66$, $\rho<.51$), which suggests that it scaled well and might have pushed participants towards an efficiency ceiling.

In terms of the number of correct answers versus ground truth, participants who used \edax (M:$3.72$, SD:$0.06$) were \emph{\textbf{2.2}} times more accurate compared to those using \pp(M:$1.70$, SD:3.56, t(29)=$2.791$, $\rho<.001$), which suggests that \edax  better assisted users in analyzing and reduced the risk of false discoveries. We again found that there was no significant difference detected between datasets (t(14)=$.4156$, $\rho<.1299$), however, as in completed tasks, we find that \pp did a significantly better job for small datasets and failed to guide users for larger ones (t(14)=$-1.27$, $\rho<.00042$). These results are encouraging: \edax can help participants with different skill-levels complete many tasks with fewer errors.

\begin{figure}[t]
    \centering
    \includegraphics[width=0.42\textwidth]{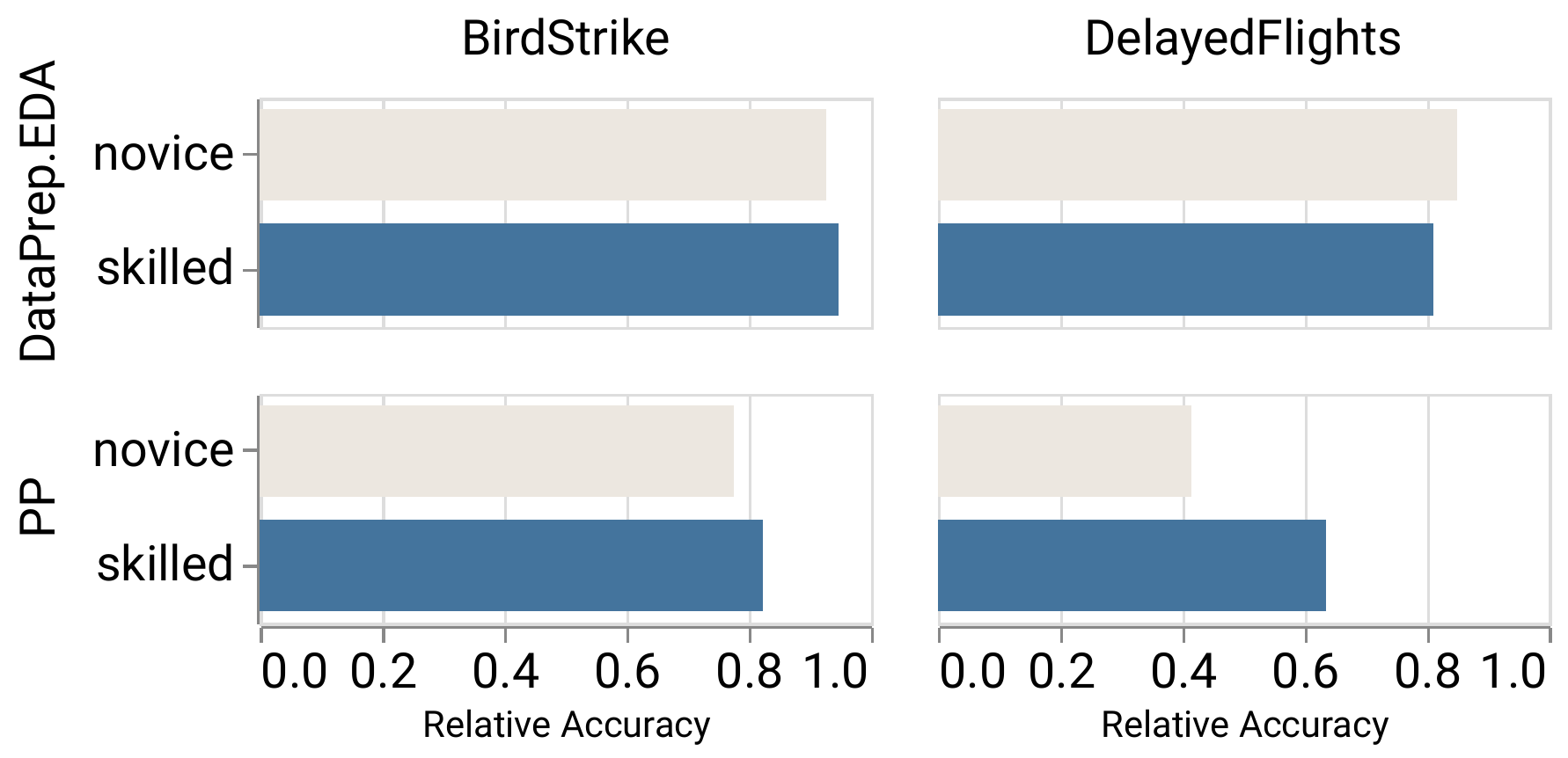}
    \vspace{-1em}
    \caption{\add{Relative Accuracy of \edax and \pp across different skill levels of participants in dataset BirdStrike and DelayedFlights.}}
    \vspace{-1em}
    \label{fig:comparison_groups_accuracy}
    \vspace{-.5em}
\end{figure}

As the number of completed tasks affects the amount of (in)correct answers, we used our relative accuracy metric. The average relative accuracy of \edax (M: .82 , SD: 0.07) among participants was \emph{\textbf{1.5}} times higher than \pp (M: .53 , SD: 1.35). Considering expertise and dataset complexity (Figure \ref{fig:comparison_groups_accuracy}), we find that users from both skill levels achieve similar relative accuracy in both datasets, but skilled participants did significantly better than novice participants only for \pp in complex datasets. This suggests that \edax performed better at leveling skill differences and dataset complexity. 

\stitle{Qualitative feedback.} \add{In our post-survey we also asked participants to share comments and feedback to add context to the performance differences we observed. In our quantitative results, participants often referenced the responsiveness and efficiency of \edax (“fast and responsive”) and took issue with the speed of Pandas-profiling (“it didn't work, took forever to process”, "I would also like to use \pp if the efficiency is not the bottleneck"). We also asked how more granular information affected their performance. Participants reflected that they felt more control (``I can find the necessary information very quickly, and that really helps a lot for me to solve the problems and questions very quickly", “felt like I had more control, simpler results”) and accessible (“I find all the answers I need, and \edax is more easy to understand”).}

\msection{Conclusion \& Future Work}\label{sec:conclusion}

In this paper, we proposed a task-centric EDA tool design pattern and built such a system in Python, called \edax. We carefully designed the API in \edax, mapping common EDA tasks in statistical modeling to corresponding stats/plots. 
Additionally, \edax provides tools for automatically generating insights and providing guides. 
Through our implementation, we discussed several issues in building data processing pipelines using \dask and presented our solutions. We conducted a performance evaluation on 15 real-world data science datasets from Kaggle and a user study with 32 participants. The results showed that \edax significantly outperformed \pp in terms of both speed and user experience.

We believe that task-centric EDA is a promising research direction. There are many interesting research problems to explore in the future. Firstly, there are some other EDA tasks for statistical modeling. For example, time-series analysis is a common EDA task in finance (e.g., stock price analysis). It would be interesting to study how to design a task-centric API for these tasks as well. Secondly, we notice that the speedup of \edax over \pandas tends to get small when IO becomes the bottleneck. We plan to investigate how to reduce IO cost using data compression techniques and column store.  \add{Thirdly, we plan to leverage sampling and sketches to speed up computation. The challenges are i) how to detect the scenarios of applying sampling/sketches; ii) how to notify users of the possible risk of sampling/sketches in a user-friendly way.}

\clearpage
\bibliographystyle{ACM-Reference-Format}
\bibliography{ref}


\begin{thebibliography}{73}


\ifx \showCODEN    \undefined \def \showCODEN     #1{\unskip}     \fi
\ifx \showDOI      \undefined \def \showDOI       #1{#1}\fi
\ifx \showISBNx    \undefined \def \showISBNx     #1{\unskip}     \fi
\ifx \showISBNxiii \undefined \def \showISBNxiii  #1{\unskip}     \fi
\ifx \showISSN     \undefined \def \showISSN      #1{\unskip}     \fi
\ifx \showLCCN     \undefined \def \showLCCN      #1{\unskip}     \fi
\ifx \shownote     \undefined \def \shownote      #1{#1}          \fi
\ifx \showarticletitle \undefined \def \showarticletitle #1{#1}   \fi
\ifx \showURL      \undefined \def \showURL       {\relax}        \fi
\providecommand\bibfield[2]{#2}
\providecommand\bibinfo[2]{#2}
\providecommand\natexlab[1]{#1}
\providecommand\showeprint[2][]{arXiv:#2}

\bibitem[\protect\citeauthoryear{??}{dat}{2020a}]%
        {dat:conflicts}
 \bibinfo{year}{2020}\natexlab{a}.
\newblock \bibinfo{title}{{ACLED Asian Conflicts, 2015-2017}}.
\newblock
\newblock
\urldef\tempurl%
\url{https://www.kaggle.com/jboysen/asian-conflicts}
\showURL{%
Retrieved September 22, 2020 from \tempurl}


\bibitem[\protect\citeauthoryear{??}{dat}{2020b}]%
        {dat:adult}
 \bibinfo{year}{2020}\natexlab{b}.
\newblock \bibinfo{title}{{Adult Census Income}}.
\newblock
\newblock
\urldef\tempurl%
\url{https://www.kaggle.com/uciml/adult-census-income}
\showURL{%
Retrieved September 22, 2020 from \tempurl}


\bibitem[\protect\citeauthoryear{??}{alt}{2020}]%
        {alteryx}
 \bibinfo{year}{2020}\natexlab{}.
\newblock \bibinfo{title}{{Alteryx: Automation that lets data speak and people
  think}}.
\newblock
\newblock
\urldef\tempurl%
\url{https://www.alteryx.com/}
\showURL{%
Retrieved September 22, 2020 from \tempurl}


\bibitem[\protect\citeauthoryear{??}{dat}{2020c}]%
        {dat:automobile}
 \bibinfo{year}{2020}\natexlab{c}.
\newblock \bibinfo{title}{{Automobile Dataset}}.
\newblock
\newblock
\urldef\tempurl%
\url{https://www.kaggle.com/toramky/automobile-dataset}
\showURL{%
Retrieved September 22, 2020 from \tempurl}


\bibitem[\protect\citeauthoryear{??}{aut}{2020}]%
        {autoviz}
 \bibinfo{year}{2020}\natexlab{}.
\newblock \bibinfo{title}{{AutoViz: Automatically Visualize any dataset, any
  size with a single line of code}}.
\newblock
\newblock
\urldef\tempurl%
\url{https://github.com/AutoViML/AutoViz}
\showURL{%
Retrieved September 22, 2020 from \tempurl}


\bibitem[\protect\citeauthoryear{??}{dat}{2020d}]%
        {dat:basketball}
 \bibinfo{year}{2020}\natexlab{d}.
\newblock \bibinfo{title}{{Basketball Players Stats per Season - 49 Leagues}}.
\newblock
\newblock
\urldef\tempurl%
\url{https://www.kaggle.com/jacobbaruch/basketball-players-stats-per-season-49-leagues}
\showURL{%
Retrieved September 22, 2020 from \tempurl}


\bibitem[\protect\citeauthoryear{??}{dat}{2020e}]%
        {dat:bitcoin}
 \bibinfo{year}{2020}\natexlab{e}.
\newblock \bibinfo{title}{{Bitcoin Dataset}}.
\newblock
\newblock
\urldef\tempurl%
\url{https://www.kaggle.com/mczielinski/bitcoin-historical-data}
\showURL{%
Retrieved September 22, 2020 from \tempurl}


\bibitem[\protect\citeauthoryear{??}{dat}{2020f}]%
        {dat:chess}
 \bibinfo{year}{2020}\natexlab{f}.
\newblock \bibinfo{title}{{Chess Game Dataset (Lichess)}}.
\newblock
\newblock
\urldef\tempurl%
\url{https://www.kaggle.com/datasnaek/chess}
\showURL{%
Retrieved September 22, 2020 from \tempurl}


\bibitem[\protect\citeauthoryear{??}{edx}{2020}]%
        {edx-ds}
 \bibinfo{year}{2020}\natexlab{}.
\newblock \bibinfo{title}{{Data science courses on edX}}.
\newblock
\newblock
\urldef\tempurl%
\url{https://www.edx.org/course/subject/data-science#python}
\showURL{%
Retrieved September 22, 2020 from \tempurl}


\bibitem[\protect\citeauthoryear{??}{dat}{2020g}]%
        {dataexplorer}
 \bibinfo{year}{2020}\natexlab{g}.
\newblock \bibinfo{title}{{DataExplorer: Automate Data Exploration and
  Treatment}}.
\newblock
\newblock
\urldef\tempurl%
\url{https://cran.r-project.org/web/packages/DataExplorer/vignettes/dataexplorer-intro.html}
\showURL{%
Retrieved September 22, 2020 from \tempurl}


\bibitem[\protect\citeauthoryear{??}{dat}{2020h}]%
        {dat:credit}
 \bibinfo{year}{2020}\natexlab{h}.
\newblock \bibinfo{title}{{Default of Credit Card Clients Dataset}}.
\newblock
\newblock
\urldef\tempurl%
\url{https://www.kaggle.com/uciml/default-of-credit-card-clients-dataset}
\showURL{%
Retrieved September 22, 2020 from \tempurl}


\bibitem[\protect\citeauthoryear{??}{dat}{2020i}]%
        {dat:diamonds}
 \bibinfo{year}{2020}\natexlab{i}.
\newblock \bibinfo{title}{{Diamonds}}.
\newblock
\newblock
\urldef\tempurl%
\url{https://www.kaggle.com/shivam2503/diamonds}
\showURL{%
Retrieved September 22, 2020 from \tempurl}


\bibitem[\protect\citeauthoryear{??}{dat}{2020j}]%
        {data-to-viz}
 \bibinfo{year}{2020}\natexlab{j}.
\newblock \bibinfo{title}{{From Data to Viz}}.
\newblock
\newblock
\urldef\tempurl%
\url{https://www.data-to-viz.com/}
\showURL{%
Retrieved September 22, 2020 from \tempurl}


\bibitem[\protect\citeauthoryear{??}{dat}{2020k}]%
        {dat:heart}
 \bibinfo{year}{2020}\natexlab{k}.
\newblock \bibinfo{title}{{Heart Disease UCI}}.
\newblock
\newblock
\urldef\tempurl%
\url{https://www.kaggle.com/ronitf/heart-disease-uci}
\showURL{%
Retrieved September 22, 2020 from \tempurl}


\bibitem[\protect\citeauthoryear{??}{dat}{2020l}]%
        {dat:hotel}
 \bibinfo{year}{2020}\natexlab{l}.
\newblock \bibinfo{title}{{Hotel booking demand}}.
\newblock
\newblock
\urldef\tempurl%
\url{https://www.kaggle.com/jessemostipak/hotel-booking-demand}
\showURL{%
Retrieved September 22, 2020 from \tempurl}


\bibitem[\protect\citeauthoryear{??}{ibm}{2020}]%
        {ibm-ds}
 \bibinfo{year}{2020}\natexlab{}.
\newblock \bibinfo{title}{{IBM Data Science Professional Certificate}}.
\newblock
\newblock
\urldef\tempurl%
\url{https://www.coursera.org/professional-certificates/ibm-data-science}
\showURL{%
Retrieved September 22, 2020 from \tempurl}


\bibitem[\protect\citeauthoryear{??}{sps}{2020}]%
        {spss}
 \bibinfo{year}{2020}\natexlab{}.
\newblock \bibinfo{title}{{IBM SPSS Statistics: Easy-to-Use Data Analysis}}.
\newblock
\newblock
\urldef\tempurl%
\url{https://www.ibm.com/analytics/spss-statistics-software}
\showURL{%
Retrieved September 22, 2020 from \tempurl}


\bibitem[\protect\citeauthoryear{??}{jmp}{2020}]%
        {jmp}
 \bibinfo{year}{2020}\natexlab{}.
\newblock \bibinfo{title}{{JMP: Statistical Discovery From SAS}}.
\newblock
\newblock
\urldef\tempurl%
\url{https://www.jmp.com}
\showURL{%
Retrieved September 22, 2020 from \tempurl}


\bibitem[\protect\citeauthoryear{??}{kag}{2020}]%
        {kaggle}
 \bibinfo{year}{2020}\natexlab{}.
\newblock \bibinfo{title}{{Kaggle: Your Machine Learning and Data Science
  Community}}.
\newblock
\newblock
\urldef\tempurl%
\url{https://www.kaggle.com/}
\showURL{%
Retrieved September 22, 2020 from \tempurl}


\bibitem[\protect\citeauthoryear{??}{lux}{2020}]%
        {lux}
 \bibinfo{year}{2020}\natexlab{}.
\newblock \bibinfo{title}{{Lux: A Python API for Intelligent Visual
  Discovery}}.
\newblock
\newblock
\urldef\tempurl%
\url{https://github.com/lux-org/lux}
\showURL{%
Retrieved September 22, 2020 from \tempurl}


\bibitem[\protect\citeauthoryear{??}{exc}{2020}]%
        {excel}
 \bibinfo{year}{2020}\natexlab{}.
\newblock \bibinfo{title}{{Microsoft Excel: Work together on Excel
  spreadsheets}}.
\newblock
\newblock
\urldef\tempurl%
\url{https://www.microsoft.com/en-us/microsoft-365/excel}
\showURL{%
Retrieved September 22, 2020 from \tempurl}


\bibitem[\protect\citeauthoryear{??}{pow}{2020}]%
        {powerbi}
 \bibinfo{year}{2020}\natexlab{}.
\newblock \bibinfo{title}{{Microsoft Power BI: Data Visualization}}.
\newblock
\newblock
\urldef\tempurl%
\url{https://powerbi.microsoft.com/en-us/}
\showURL{%
Retrieved September 22, 2020 from \tempurl}


\bibitem[\protect\citeauthoryear{??}{dat}{2020m}]%
        {dat:diabetes}
 \bibinfo{year}{2020}\natexlab{m}.
\newblock \bibinfo{title}{{Pima Indians Diabetes Database}}.
\newblock
\newblock
\urldef\tempurl%
\url{https://www.kaggle.com/uciml/pima-indians-diabetes-database}
\showURL{%
Retrieved September 22, 2020 from \tempurl}


\bibitem[\protect\citeauthoryear{??}{ude}{2020}]%
        {udemy-ds}
 \bibinfo{year}{2020}\natexlab{}.
\newblock \bibinfo{title}{{Python for Data Science and Machine Learning
  Bootcamp}}.
\newblock
\newblock
\urldef\tempurl%
\url{https://www.udemy.com/course/python-for-data-science-and-machine-learning-bootcamp/}
\showURL{%
Retrieved September 22, 2020 from \tempurl}


\bibitem[\protect\citeauthoryear{??}{qli}{2020}]%
        {qlik}
 \bibinfo{year}{2020}\natexlab{}.
\newblock \bibinfo{title}{{Qlik: Data Analytics and Data Integration
  Solutions}}.
\newblock
\newblock
\urldef\tempurl%
\url{https://www.qlik.com/us/}
\showURL{%
Retrieved September 22, 2020 from \tempurl}


\bibitem[\protect\citeauthoryear{??}{dat}{2020n}]%
        {dat:rain}
 \bibinfo{year}{2020}\natexlab{n}.
\newblock \bibinfo{title}{{Rain in Australia}}.
\newblock
\newblock
\urldef\tempurl%
\url{https://www.kaggle.com/jsphyg/weather-dataset-rattle-package}
\showURL{%
Retrieved September 22, 2020 from \tempurl}


\bibitem[\protect\citeauthoryear{??}{sas}{2020}]%
        {sas}
 \bibinfo{year}{2020}\natexlab{}.
\newblock \bibinfo{title}{{SAS: Analytics, Artificial Intelligence and Data
  Management}}.
\newblock
\newblock
\urldef\tempurl%
\url{https://www.sas.com/}
\showURL{%
Retrieved September 22, 2020 from \tempurl}


\bibitem[\protect\citeauthoryear{??}{dat}{2020o}]%
        {dat:solar}
 \bibinfo{year}{2020}\natexlab{o}.
\newblock \bibinfo{title}{{Solar Radiation Prediction}}.
\newblock
\newblock
\urldef\tempurl%
\url{https://www.kaggle.com/dronio/SolarEnergy}
\showURL{%
Retrieved September 22, 2020 from \tempurl}


\bibitem[\protect\citeauthoryear{??}{spl}{2020}]%
        {splunk}
 \bibinfo{year}{2020}\natexlab{}.
\newblock \bibinfo{title}{{splunk: The Data-to-Everything Platform}}.
\newblock
\newblock
\urldef\tempurl%
\url{https://www.splunk.com/}
\showURL{%
Retrieved September 22, 2020 from \tempurl}


\bibitem[\protect\citeauthoryear{??}{dat}{2020p}]%
        {dat:suicide}
 \bibinfo{year}{2020}\natexlab{p}.
\newblock \bibinfo{title}{{Suicide Rates Overview 1985 to 2016}}.
\newblock
\newblock
\urldef\tempurl%
\url{https://www.kaggle.com/russellyates88/suicide-rates-overview-1985-to-2016}
\showURL{%
Retrieved September 22, 2020 from \tempurl}


\bibitem[\protect\citeauthoryear{??}{swe}{2020}]%
        {sweetviz}
 \bibinfo{year}{2020}\natexlab{}.
\newblock \bibinfo{title}{{Sweetviz: an open source Python library that
  generates beautiful, high-density visualizations to kickstart EDA
  (Exploratory Data Analysis) with a single line of code}}.
\newblock
\newblock
\urldef\tempurl%
\url{https://github.com/fbdesignpro/sweetviz}
\showURL{%
Retrieved September 22, 2020 from \tempurl}


\bibitem[\protect\citeauthoryear{??}{tab}{2020}]%
        {tableau}
 \bibinfo{year}{2020}\natexlab{}.
\newblock \bibinfo{title}{{Tableau: an interactive data visualization software
  company}}.
\newblock
\newblock
\urldef\tempurl%
\url{https://www.tableau.com/}
\showURL{%
Retrieved September 22, 2020 from \tempurl}


\bibitem[\protect\citeauthoryear{??}{tio}{2020}]%
        {tiobe-index}
 \bibinfo{year}{2020}\natexlab{}.
\newblock \bibinfo{title}{{The TIOBE Programming Community Index}}.
\newblock
\newblock
\urldef\tempurl%
\url{https://www.tiobe.com/tiobe-index}
\showURL{%
Retrieved September 22, 2020 from \tempurl}


\bibitem[\protect\citeauthoryear{??}{ucb}{2020}]%
        {ucb-ds}
 \bibinfo{year}{2020}\natexlab{}.
\newblock \bibinfo{title}{{The UC Berkeley Foundations of Data Science
  Course}}.
\newblock
\newblock
\urldef\tempurl%
\url{http://data8.org/}
\showURL{%
Retrieved September 22, 2020 from \tempurl}


\bibitem[\protect\citeauthoryear{??}{spo}{2020}]%
        {spotfire}
 \bibinfo{year}{2020}\natexlab{}.
\newblock \bibinfo{title}{{TIBCO Spotfire Data Visualization and Analytics
  Software}}.
\newblock
\newblock
\urldef\tempurl%
\url{https://www.tibco.com/products/tibco-spotfire}
\showURL{%
Retrieved September 22, 2020 from \tempurl}


\bibitem[\protect\citeauthoryear{??}{dat}{2020q}]%
        {dat:titanic}
 \bibinfo{year}{2020}\natexlab{q}.
\newblock \bibinfo{title}{{Titanic: Machine Learning from Disaster}}.
\newblock
\newblock
\urldef\tempurl%
\url{https://www.kaggle.com/c/titanic/data?select=train.csv}
\showURL{%
Retrieved September 22, 2020 from \tempurl}


\bibitem[\protect\citeauthoryear{??}{dat}{2020r}]%
        {dat:women}
 \bibinfo{year}{2020}\natexlab{r}.
\newblock \bibinfo{title}{{Top Women Chess Players}}.
\newblock
\newblock
\urldef\tempurl%
\url{https://www.kaggle.com/vikasojha98/top-women-chess-players}
\showURL{%
Retrieved September 22, 2020 from \tempurl}


\bibitem[\protect\citeauthoryear{??}{koa}{2021}]%
        {koalas}
 \bibinfo{year}{2021}\natexlab{}.
\newblock \bibinfo{title}{{Koalas: pandas API on Apache Spark}}.
\newblock
\newblock
\urldef\tempurl%
\url{https://github.com/databricks/koalas}
\showURL{%
Retrieved February 9, 2021 from \tempurl}


\bibitem[\protect\citeauthoryear{Abedjan, Golab, and Naumann}{Abedjan
  et~al\mbox{.}}{2015}]%
        {abedjan2015profiling}
\bibfield{author}{\bibinfo{person}{Ziawasch Abedjan}, \bibinfo{person}{Lukasz
  Golab}, {and} \bibinfo{person}{Felix Naumann}.}
  \bibinfo{year}{2015}\natexlab{}.
\newblock \showarticletitle{Profiling relational data: a survey}.
\newblock \bibinfo{journal}{\emph{The VLDB Journal}} \bibinfo{volume}{24},
  \bibinfo{number}{4} (\bibinfo{year}{2015}), \bibinfo{pages}{557--581}.
\newblock


\bibitem[\protect\citeauthoryear{Battle and Heer}{Battle and Heer}{2019}]%
        {battle2019characterizing}
\bibfield{author}{\bibinfo{person}{Leilani Battle} {and}
  \bibinfo{person}{Jeffrey Heer}.} \bibinfo{year}{2019}\natexlab{}.
\newblock \showarticletitle{Characterizing exploratory visual analysis: A
  literature review and evaluation of analytic provenance in tableau}. In
  \bibinfo{booktitle}{\emph{Computer Graphics Forum}},
  Vol.~\bibinfo{volume}{38}. Wiley Online Library, \bibinfo{pages}{145--159}.
\newblock


\bibitem[\protect\citeauthoryear{Bilogur}{Bilogur}{2018}]%
        {bilogur2018missingno}
\bibfield{author}{\bibinfo{person}{Aleksey Bilogur}.}
  \bibinfo{year}{2018}\natexlab{}.
\newblock \showarticletitle{Missingno: a missing data visualization suite}.
\newblock \bibinfo{journal}{\emph{Journal of Open Source Software}}
  \bibinfo{volume}{3}, \bibinfo{number}{22} (\bibinfo{year}{2018}),
  \bibinfo{pages}{547}.
\newblock


\bibitem[\protect\citeauthoryear{{Bokeh Development Team}}{{Bokeh Development
  Team}}{2018}]%
        {bokeh}
\bibfield{author}{\bibinfo{person}{{Bokeh Development Team}}.}
  \bibinfo{year}{2018}\natexlab{}.
\newblock \bibinfo{booktitle}{\emph{Bokeh: Python library for interactive
  visualization}}.
\newblock
\urldef\tempurl%
\url{https://bokeh.pydata.org/en/latest/}
\showURL{%
\tempurl}


\bibitem[\protect\citeauthoryear{Brugman}{Brugman}{2019}]%
        {pandasprofiling2019}
\bibfield{author}{\bibinfo{person}{Simon Brugman}.}
  \bibinfo{year}{2019}\natexlab{}.
\newblock \bibinfo{title}{{Pandas-profiling: Exploratory Data Analysis for
  Python}}.
\newblock
  \bibinfo{howpublished}{\url{https://github.com/pandas-profiling/pandas-profiling}}.
\newblock


\bibitem[\protect\citeauthoryear{Buitinck, Louppe, Blondel, Pedregosa, Mueller,
  Grisel, Niculae, Prettenhofer, Gramfort, Grobler, et~al\mbox{.}}{Buitinck
  et~al\mbox{.}}{2013}]%
        {buitinck2013api}
\bibfield{author}{\bibinfo{person}{Lars Buitinck}, \bibinfo{person}{Gilles
  Louppe}, \bibinfo{person}{Mathieu Blondel}, \bibinfo{person}{Fabian
  Pedregosa}, \bibinfo{person}{Andreas Mueller}, \bibinfo{person}{Olivier
  Grisel}, \bibinfo{person}{Vlad Niculae}, \bibinfo{person}{Peter
  Prettenhofer}, \bibinfo{person}{Alexandre Gramfort}, \bibinfo{person}{Jaques
  Grobler}, {et~al\mbox{.}}} \bibinfo{year}{2013}\natexlab{}.
\newblock \showarticletitle{API design for machine learning software:
  experiences from the scikit-learn project}.
\newblock \bibinfo{journal}{\emph{arXiv preprint arXiv:1309.0238}}
  (\bibinfo{year}{2013}).
\newblock


\bibitem[\protect\citeauthoryear{Card and Mackinlay}{Card and
  Mackinlay}{1997}]%
        {card1997structure}
\bibfield{author}{\bibinfo{person}{Stuart~K Card} {and} \bibinfo{person}{Jock
  Mackinlay}.} \bibinfo{year}{1997}\natexlab{}.
\newblock \showarticletitle{The structure of the information visualization
  design space}. In \bibinfo{booktitle}{\emph{Proceedings of VIZ'97:
  Visualization Conference, Information Visualization Symposium and Parallel
  Rendering Symposium}}. IEEE, \bibinfo{pages}{92--99}.
\newblock


\bibitem[\protect\citeauthoryear{Cui, Badam, Yal{\c{c}}in, and Elmqvist}{Cui
  et~al\mbox{.}}{2019}]%
        {cui2019datasite}
\bibfield{author}{\bibinfo{person}{Zhe Cui}, \bibinfo{person}{Sriram~Karthik
  Badam}, \bibinfo{person}{M~Adil Yal{\c{c}}in}, {and} \bibinfo{person}{Niklas
  Elmqvist}.} \bibinfo{year}{2019}\natexlab{}.
\newblock \showarticletitle{Datasite: Proactive visual data exploration with
  computation of insight-based recommendations}.
\newblock \bibinfo{journal}{\emph{Information Visualization}}
  \bibinfo{volume}{18}, \bibinfo{number}{2} (\bibinfo{year}{2019}),
  \bibinfo{pages}{251--267}.
\newblock


\bibitem[\protect\citeauthoryear{Demiralp, Haas, Parthasarathy, and
  Pedapati}{Demiralp et~al\mbox{.}}{2017}]%
        {demiralp2017foresight}
\bibfield{author}{\bibinfo{person}{{\c{C}}a{\u{g}}atay Demiralp},
  \bibinfo{person}{Peter~J Haas}, \bibinfo{person}{Srinivasan Parthasarathy},
  {and} \bibinfo{person}{Tejaswini Pedapati}.} \bibinfo{year}{2017}\natexlab{}.
\newblock \showarticletitle{Foresight: Recommending visual insights}.
\newblock \bibinfo{journal}{\emph{arXiv preprint arXiv:1707.03877}}
  (\bibinfo{year}{2017}).
\newblock


\bibitem[\protect\citeauthoryear{Deutch, Gilad, Milo, and Somech}{Deutch
  et~al\mbox{.}}{2020}]%
        {deutch2020explained}
\bibfield{author}{\bibinfo{person}{Daniel Deutch}, \bibinfo{person}{Amir
  Gilad}, \bibinfo{person}{Tova Milo}, {and} \bibinfo{person}{Amit Somech}.}
  \bibinfo{year}{2020}\natexlab{}.
\newblock \showarticletitle{ExplainED: explanations for EDA notebooks}.
\newblock \bibinfo{journal}{\emph{Proceedings of the VLDB Endowment}}
  \bibinfo{volume}{13}, \bibinfo{number}{12} (\bibinfo{year}{2020}),
  \bibinfo{pages}{2917--2920}.
\newblock


\bibitem[\protect\citeauthoryear{Dibia and Demiralp}{Dibia and
  Demiralp}{2019}]%
        {dibia2019data2vis}
\bibfield{author}{\bibinfo{person}{Victor Dibia} {and}
  \bibinfo{person}{{\c{C}}a{\u{g}}atay Demiralp}.}
  \bibinfo{year}{2019}\natexlab{}.
\newblock \showarticletitle{Data2vis: Automatic generation of data
  visualizations using sequence-to-sequence recurrent neural networks}.
\newblock \bibinfo{journal}{\emph{IEEE computer graphics and applications}}
  \bibinfo{volume}{39}, \bibinfo{number}{5} (\bibinfo{year}{2019}),
  \bibinfo{pages}{33--46}.
\newblock


\bibitem[\protect\citeauthoryear{Ding, Han, Xu, Zhang, and Zhang}{Ding
  et~al\mbox{.}}{2019}]%
        {ding2019quickinsights}
\bibfield{author}{\bibinfo{person}{Rui Ding}, \bibinfo{person}{Shi Han},
  \bibinfo{person}{Yong Xu}, \bibinfo{person}{Haidong Zhang}, {and}
  \bibinfo{person}{Dongmei Zhang}.} \bibinfo{year}{2019}\natexlab{}.
\newblock \showarticletitle{Quickinsights: Quick and automatic discovery of
  insights from multi-dimensional data}. In
  \bibinfo{booktitle}{\emph{Proceedings of the 2019 International Conference on
  Management of Data}}. \bibinfo{pages}{317--332}.
\newblock


\bibitem[\protect\citeauthoryear{Hu, Bakker, Li, Kraska, and Hidalgo}{Hu
  et~al\mbox{.}}{2019}]%
        {hu2019vizml}
\bibfield{author}{\bibinfo{person}{Kevin Hu}, \bibinfo{person}{Michiel~A
  Bakker}, \bibinfo{person}{Stephen Li}, \bibinfo{person}{Tim Kraska}, {and}
  \bibinfo{person}{C{\'e}sar Hidalgo}.} \bibinfo{year}{2019}\natexlab{}.
\newblock \showarticletitle{Vizml: A machine learning approach to visualization
  recommendation}. In \bibinfo{booktitle}{\emph{Proceedings of the 2019 CHI
  Conference on Human Factors in Computing Systems}}. \bibinfo{pages}{1--12}.
\newblock


\bibitem[\protect\citeauthoryear{Hunter}{Hunter}{2007}]%
        {Hunter:2007}
\bibfield{author}{\bibinfo{person}{J.~D. Hunter}.}
  \bibinfo{year}{2007}\natexlab{}.
\newblock \showarticletitle{Matplotlib: A 2D graphics environment}.
\newblock \bibinfo{journal}{\emph{Computing in Science \& Engineering}}
  \bibinfo{volume}{9}, \bibinfo{number}{3} (\bibinfo{year}{2007}),
  \bibinfo{pages}{90--95}.
\newblock
\urldef\tempurl%
\url{https://doi.org/10.1109/MCSE.2007.55}
\showDOI{\tempurl}


\bibitem[\protect\citeauthoryear{Kandel, Parikh, Paepcke, Hellerstein, and
  Heer}{Kandel et~al\mbox{.}}{2012}]%
        {kandel2012profiler}
\bibfield{author}{\bibinfo{person}{Sean Kandel}, \bibinfo{person}{Ravi Parikh},
  \bibinfo{person}{Andreas Paepcke}, \bibinfo{person}{Joseph~M Hellerstein},
  {and} \bibinfo{person}{Jeffrey Heer}.} \bibinfo{year}{2012}\natexlab{}.
\newblock \showarticletitle{Profiler: Integrated statistical analysis and
  visualization for data quality assessment}. In
  \bibinfo{booktitle}{\emph{Proceedings of the International Working Conference
  on Advanced Visual Interfaces}}. \bibinfo{pages}{547--554}.
\newblock


\bibitem[\protect\citeauthoryear{Lin, Ke, Lou, Zhang, Sui, Xu, Zhou, Qiao, and
  Zhang}{Lin et~al\mbox{.}}{2018}]%
        {lin2018bigin4}
\bibfield{author}{\bibinfo{person}{Qingwei Lin}, \bibinfo{person}{Weichen Ke},
  \bibinfo{person}{Jian-Guang Lou}, \bibinfo{person}{Hongyu Zhang},
  \bibinfo{person}{Kaixin Sui}, \bibinfo{person}{Yong Xu},
  \bibinfo{person}{Ziyi Zhou}, \bibinfo{person}{Bo Qiao}, {and}
  \bibinfo{person}{Dongmei Zhang}.} \bibinfo{year}{2018}\natexlab{}.
\newblock \showarticletitle{BigIN4: Instant, Interactive Insight Identification
  for Multi-Dimensional Big Data}. In \bibinfo{booktitle}{\emph{Proceedings of
  the 24th ACM SIGKDD International Conference on Knowledge Discovery \& Data
  Mining}}. \bibinfo{pages}{547--555}.
\newblock


\bibitem[\protect\citeauthoryear{Luo, Qin, Tang, and Li}{Luo
  et~al\mbox{.}}{2018}]%
        {luo2018deepeye}
\bibfield{author}{\bibinfo{person}{Yuyu Luo}, \bibinfo{person}{Xuedi Qin},
  \bibinfo{person}{Nan Tang}, {and} \bibinfo{person}{Guoliang Li}.}
  \bibinfo{year}{2018}\natexlab{}.
\newblock \showarticletitle{Deepeye: Towards automatic data visualization}. In
  \bibinfo{booktitle}{\emph{2018 IEEE 34th International Conference on Data
  Engineering (ICDE)}}. IEEE, \bibinfo{pages}{101--112}.
\newblock


\bibitem[\protect\citeauthoryear{Mackinlay, Hanrahan, and Stolte}{Mackinlay
  et~al\mbox{.}}{2007}]%
        {mackinlay2007show}
\bibfield{author}{\bibinfo{person}{Jock Mackinlay}, \bibinfo{person}{Pat
  Hanrahan}, {and} \bibinfo{person}{Chris Stolte}.}
  \bibinfo{year}{2007}\natexlab{}.
\newblock \showarticletitle{Show me: Automatic presentation for visual
  analysis}.
\newblock \bibinfo{journal}{\emph{IEEE transactions on visualization and
  computer graphics}} \bibinfo{volume}{13}, \bibinfo{number}{6}
  (\bibinfo{year}{2007}), \bibinfo{pages}{1137--1144}.
\newblock


\bibitem[\protect\citeauthoryear{Moritz, Wang, Nelson, Lin, Smith, Howe, and
  Heer}{Moritz et~al\mbox{.}}{2019}]%
        {2019-draco}
\bibfield{author}{\bibinfo{person}{Dominik Moritz}, \bibinfo{person}{Chenglong
  Wang}, \bibinfo{person}{Gregory Nelson}, \bibinfo{person}{Halden Lin},
  \bibinfo{person}{Adam~M. Smith}, \bibinfo{person}{Bill Howe}, {and}
  \bibinfo{person}{Jeffrey Heer}.} \bibinfo{year}{2019}\natexlab{}.
\newblock \showarticletitle{Formalizing Visualization Design Knowledge as
  Constraints: Actionable and Extensible Models in Draco}.
\newblock \bibinfo{journal}{\emph{IEEE Trans. Visualization \& Comp. Graphics
  (Proc. InfoVis)}} (\bibinfo{year}{2019}).
\newblock
\urldef\tempurl%
\url{http://idl.cs.washington.edu/papers/draco}
\showURL{%
\tempurl}


\bibitem[\protect\citeauthoryear{Papenbrock, Bergmann, Finke, Zwiener, and
  Naumann}{Papenbrock et~al\mbox{.}}{2015}]%
        {papenbrock2015data}
\bibfield{author}{\bibinfo{person}{Thorsten Papenbrock}, \bibinfo{person}{Tanja
  Bergmann}, \bibinfo{person}{Moritz Finke}, \bibinfo{person}{Jakob Zwiener},
  {and} \bibinfo{person}{Felix Naumann}.} \bibinfo{year}{2015}\natexlab{}.
\newblock \showarticletitle{Data profiling with metanome}.
\newblock \bibinfo{journal}{\emph{Proceedings of the VLDB Endowment}}
  \bibinfo{volume}{8}, \bibinfo{number}{12} (\bibinfo{year}{2015}),
  \bibinfo{pages}{1860--1863}.
\newblock


\bibitem[\protect\citeauthoryear{Peng}{Peng}{2012}]%
        {peng2012exploratory}
\bibfield{author}{\bibinfo{person}{Roger Peng}.}
  \bibinfo{year}{2012}\natexlab{}.
\newblock \bibinfo{booktitle}{\emph{Exploratory data analysis with R}}.
\newblock \bibinfo{publisher}{Lulu. com}.
\newblock


\bibitem[\protect\citeauthoryear{Petersohn, Ma, Lee, Macke, Xin, Mo, Gonzalez,
  Hellerstein, Joseph, and Parameswaran}{Petersohn et~al\mbox{.}}{2020}]%
        {DBLP:journals/corr/abs-2001-00888}
\bibfield{author}{\bibinfo{person}{Devin Petersohn},
  \bibinfo{person}{William~W. Ma}, \bibinfo{person}{Doris Jung~Lin Lee},
  \bibinfo{person}{Stephen Macke}, \bibinfo{person}{Doris Xin},
  \bibinfo{person}{Xiangxi Mo}, \bibinfo{person}{Joseph~E. Gonzalez},
  \bibinfo{person}{Joseph~M. Hellerstein}, \bibinfo{person}{Anthony~D. Joseph},
  {and} \bibinfo{person}{Aditya~G. Parameswaran}.}
  \bibinfo{year}{2020}\natexlab{}.
\newblock \showarticletitle{Towards Scalable Dataframe Systems}.
\newblock \bibinfo{journal}{\emph{CoRR}}  \bibinfo{volume}{abs/2001.00888}
  (\bibinfo{year}{2020}).
\newblock
\showeprint[arxiv]{2001.00888}
\urldef\tempurl%
\url{http://arxiv.org/abs/2001.00888}
\showURL{%
\tempurl}


\bibitem[\protect\citeauthoryear{Raman and Hellerstein}{Raman and
  Hellerstein}{2001}]%
        {raman2001potter}
\bibfield{author}{\bibinfo{person}{Vijayshankar Raman} {and}
  \bibinfo{person}{Joseph~M Hellerstein}.} \bibinfo{year}{2001}\natexlab{}.
\newblock \showarticletitle{Potter's wheel: An interactive data cleaning
  system}. In \bibinfo{booktitle}{\emph{VLDB}}, Vol.~\bibinfo{volume}{1}.
  \bibinfo{pages}{381--390}.
\newblock


\bibitem[\protect\citeauthoryear{Seltman}{Seltman}{2012}]%
        {seltman2012experimental}
\bibfield{author}{\bibinfo{person}{Howard~J Seltman}.}
  \bibinfo{year}{2012}\natexlab{}.
\newblock \bibinfo{title}{Experimental design and analysis}.
\newblock
\newblock


\bibitem[\protect\citeauthoryear{Siddiqui, Kim, Lee, Karahalios, and
  Parameswaran}{Siddiqui et~al\mbox{.}}{2016}]%
        {siddiqui2016zenvisage}
\bibfield{author}{\bibinfo{person}{Tarique Siddiqui}, \bibinfo{person}{Albert
  Kim}, \bibinfo{person}{John Lee}, \bibinfo{person}{Karrie Karahalios}, {and}
  \bibinfo{person}{Aditya Parameswaran}.} \bibinfo{year}{2016}\natexlab{}.
\newblock \showarticletitle{zenvisage: Effortless visual data exploration}. In
  \bibinfo{booktitle}{\emph{Proceedings of the 2016 ACM SIGMOD International
  Conference on Management of Data}}.
\newblock


\bibitem[\protect\citeauthoryear{Staniak and Biecek}{Staniak and
  Biecek}{2019}]%
        {staniak2019landscape}
\bibfield{author}{\bibinfo{person}{Mateusz Staniak} {and}
  \bibinfo{person}{Przemyslaw Biecek}.} \bibinfo{year}{2019}\natexlab{}.
\newblock \showarticletitle{The Landscape of R Packages for Automated
  Exploratory Data Analysis}.
\newblock \bibinfo{journal}{\emph{arXiv preprint arXiv:1904.02101}}
  (\bibinfo{year}{2019}).
\newblock


\bibitem[\protect\citeauthoryear{Tang, Han, Yiu, Ding, and Zhang}{Tang
  et~al\mbox{.}}{2017}]%
        {tang2017topkinsights}
\bibfield{author}{\bibinfo{person}{Bo Tang}, \bibinfo{person}{Shi Han},
  \bibinfo{person}{Man Yiu}, \bibinfo{person}{Rui Ding}, {and}
  \bibinfo{person}{Dongmei Zhang}.} \bibinfo{year}{2017}\natexlab{}.
\newblock \showarticletitle{Extracting Top-K Insights from Multi-dimensional
  Data}. \bibinfo{pages}{1509--1524}.
\newblock
\urldef\tempurl%
\url{https://doi.org/10.1145/3035918.3035922}
\showDOI{\tempurl}


\bibitem[\protect\citeauthoryear{Tierney}{Tierney}{2017}]%
        {tierney2017visdat}
\bibfield{author}{\bibinfo{person}{Nicholas Tierney}.}
  \bibinfo{year}{2017}\natexlab{}.
\newblock \showarticletitle{visdat: Visualising whole data frames}.
\newblock \bibinfo{journal}{\emph{Journal of Open Source Software}}
  \bibinfo{volume}{2}, \bibinfo{number}{16} (\bibinfo{year}{2017}),
  \bibinfo{pages}{355}.
\newblock


\bibitem[\protect\citeauthoryear{Vartak, Rahman, Madden, Parameswaran, and
  Polyzotis}{Vartak et~al\mbox{.}}{2015}]%
        {vartak2015seedb}
\bibfield{author}{\bibinfo{person}{Manasi Vartak}, \bibinfo{person}{Sajjadur
  Rahman}, \bibinfo{person}{Samuel Madden}, \bibinfo{person}{Aditya
  Parameswaran}, {and} \bibinfo{person}{Neoklis Polyzotis}.}
  \bibinfo{year}{2015}\natexlab{}.
\newblock \showarticletitle{Seedb: Efficient data-driven visualization
  recommendations to support visual analytics}. In
  \bibinfo{booktitle}{\emph{Proceedings of the VLDB Endowment International
  Conference on Very Large Data Bases}}, Vol.~\bibinfo{volume}{8}. NIH Public
  Access, \bibinfo{pages}{2182}.
\newblock


\bibitem[\protect\citeauthoryear{Waskom and the seaborn~development
  team}{Waskom and the seaborn~development team}{2020}]%
        {waskom2020seaborn}
\bibfield{author}{\bibinfo{person}{Michael Waskom} {and} \bibinfo{person}{the
  seaborn~development team}.} \bibinfo{year}{2020}\natexlab{}.
\newblock \bibinfo{booktitle}{\emph{mwaskom/seaborn}}.
\newblock
\urldef\tempurl%
\url{https://doi.org/10.5281/zenodo.592845}
\showDOI{\tempurl}


\bibitem[\protect\citeauthoryear{Wickham and Grolemund}{Wickham and
  Grolemund}{2016}]%
        {wickham2016r}
\bibfield{author}{\bibinfo{person}{Hadley Wickham} {and}
  \bibinfo{person}{Garrett Grolemund}.} \bibinfo{year}{2016}\natexlab{}.
\newblock \bibinfo{booktitle}{\emph{R for data science: import, tidy,
  transform, visualize, and model data}}.
\newblock \bibinfo{publisher}{" O'Reilly Media, Inc."}.
\newblock


\bibitem[\protect\citeauthoryear{Wongsuphasawat, Moritz, Anand, Mackinlay,
  Howe, and Heer}{Wongsuphasawat et~al\mbox{.}}{2015}]%
        {wongsuphasawat2015voyager}
\bibfield{author}{\bibinfo{person}{Kanit Wongsuphasawat},
  \bibinfo{person}{Dominik Moritz}, \bibinfo{person}{Anushka Anand},
  \bibinfo{person}{Jock Mackinlay}, \bibinfo{person}{Bill Howe}, {and}
  \bibinfo{person}{Jeffrey Heer}.} \bibinfo{year}{2015}\natexlab{}.
\newblock \showarticletitle{Voyager: Exploratory analysis via faceted browsing
  of visualization recommendations}.
\newblock \bibinfo{journal}{\emph{IEEE transactions on visualization and
  computer graphics}} \bibinfo{volume}{22}, \bibinfo{number}{1}
  (\bibinfo{year}{2015}), \bibinfo{pages}{649--658}.
\newblock


\bibitem[\protect\citeauthoryear{Wongsuphasawat, Qu, Moritz, Chang, Ouk, Anand,
  Mackinlay, Howe, and Heer}{Wongsuphasawat et~al\mbox{.}}{2017}]%
        {wongsuphasawat2017voyager}
\bibfield{author}{\bibinfo{person}{Kanit Wongsuphasawat},
  \bibinfo{person}{Zening Qu}, \bibinfo{person}{Dominik Moritz},
  \bibinfo{person}{Riley Chang}, \bibinfo{person}{Felix Ouk},
  \bibinfo{person}{Anushka Anand}, \bibinfo{person}{Jock Mackinlay},
  \bibinfo{person}{Bill Howe}, {and} \bibinfo{person}{Jeffrey Heer}.}
  \bibinfo{year}{2017}\natexlab{}.
\newblock \showarticletitle{Voyager 2: Augmenting visual analysis with partial
  view specifications}. In \bibinfo{booktitle}{\emph{Proceedings of the 2017
  CHI Conference on Human Factors in Computing Systems}}.
  \bibinfo{pages}{2648--2659}.
\newblock


\bibitem[\protect\citeauthoryear{Yan, Gu, and Rzeszotarski}{Yan
  et~al\mbox{.}}{2021}]%
        {Tessera21}
\bibfield{author}{\bibinfo{person}{Jing~Nathan Yan}, \bibinfo{person}{Ziwei
  Gu}, {and} \bibinfo{person}{Jeffrey~M. Rzeszotarski}.}
  \bibinfo{year}{2021}\natexlab{}.
\newblock \showarticletitle{Tessera: Discretizing Data Analysis Workflows on a
  Task Level}. In \bibinfo{booktitle}{\emph{{CHI} '21: {CHI} Conference on
  Human Factors in Computing Systems, Yokohama, Japan, May 8--13, 2021}}.
  \bibinfo{publisher}{{ACM}}, \bibinfo{pages}{1--15}.
\newblock
\urldef\tempurl%
\url{https://doi.org/10.1145/3411764.3445728}
\showDOI{\tempurl}


\bibitem[\protect\citeauthoryear{Zaharia, Chowdhury, Franklin, Shenker, and
  Stoica}{Zaharia et~al\mbox{.}}{2010}]%
        {DBLP:conf/hotcloud/ZahariaCFSS10}
\bibfield{author}{\bibinfo{person}{Matei Zaharia}, \bibinfo{person}{Mosharaf
  Chowdhury}, \bibinfo{person}{Michael~J. Franklin}, \bibinfo{person}{Scott
  Shenker}, {and} \bibinfo{person}{Ion Stoica}.}
  \bibinfo{year}{2010}\natexlab{}.
\newblock \showarticletitle{Spark: Cluster Computing with Working Sets}. In
  \bibinfo{booktitle}{\emph{2nd {USENIX} Workshop on Hot Topics in Cloud
  Computing, HotCloud'10, Boston, MA, USA, June 22, 2010}},
  \bibfield{editor}{\bibinfo{person}{Erich~M. Nahum} {and}
  \bibinfo{person}{Dongyan Xu}} (Eds.). \bibinfo{publisher}{{USENIX}
  Association}.
\newblock
\urldef\tempurl%
\url{https://www.usenix.org/conference/hotcloud-10/spark-cluster-computing-working-sets}
\showURL{%
\tempurl}


\end{thebibliography}

\end{document}